\documentclass[%
reprint,
superscriptaddress,
 amsmath,amssymb,
 aps,
prd,
]{revtex4-2}

\usepackage{graphicx}
\usepackage{dcolumn}
\usepackage{bm}
\usepackage{lineno}
\usepackage{xcolor}
\usepackage{multirow}

\newcommand*{\MIT }{Massachusetts Institute of Technology, Cambridge, Massachusetts 02139, USA}

\newcommand*{\JLAB}{Thomas Jefferson National Accelerator Facility, Newport News, Virginia 23606, USA}
\newcommand*{\TAU }{School of Physics and Astronomy, Tel Aviv University, Tel Aviv 69978, Israel}

\newcommand*{\GW}{The George Washington University, Washington, D.C., 20052, USA}

\newcommand*{\Duke}{Duke University, Durham, North Carolina 27708, USA}
\newcommand*{\MSU}{Mississippi State University, Mississippi State, MS 39762, USA}
\newcommand*{\FIU}{Florida International University, Miami, FL 33199, USA}
\newcommand*{\UR}{University of Regina, Regina, SK S4S 0A2, Canada}
\newcommand*{\CMU}{Carnegie Mellon University, Pittsburgh, PA 15213, USA}

\begin{document}
\setlength{\extrarowheight}{2pt}


\title{Search for axion-like particles through nuclear Primakoff production using the GlueX detector}

\author{J.R.~Pybus}
\email{jrpybus@mit.edu}
\affiliation{\MIT}

\author{T.~Kolar}%
\affiliation{\TAU}%

\author{B.~Devkota}%
\affiliation{\MSU}%

\author{P.~Sharp}%
\affiliation{\GW}%

\author{B.~Yu}%
\affiliation{\Duke}%

\author{O.~Hen}%
\affiliation{\MIT}%

\author{E.~Piasetzky}%
\affiliation{\TAU}%

\author{S.N.~Santiesteban}
\affiliation{University of New Hampshire, Durham, New Hampshire 03824, USA}

\author{A.~Schmidt}%
\affiliation{\GW}%

\author{A.~Somov}%
\affiliation{\JLAB}%

\author{Y.~Soreq}%
\affiliation{Technion-Israel Institute of Technology, Haifa 32000, Israel}%

\author{H.~Szumila-Vance}%
\affiliation{\JLAB}%

\author{C.S.~Akondi}%
\affiliation{Florida State University, Tallahassee, FL 32306, USA}%

\author{C.~Ayerbe Gayoso}%
\affiliation{\MSU}%

\author{V.V.~Berdnikov}%
\affiliation{\JLAB}%

\author{H.~Bhatt}%
\affiliation{\MSU}%

\author{D.~Bhetuwal}%
\affiliation{\MSU}%

\author{M.M.~Dalton}%
\affiliation{\JLAB}%

\author{A.~Deur}%
\affiliation{\JLAB}%

\author{R.~Dotel}%
\affiliation{\FIU}%

\author{C.~Fanelli}%
\affiliation{The College of William and Mary, Williamsburg, VA 23185, USA}%

\author{J.~Guo}%
\affiliation{\CMU}%

\author{T.J.~Hague}%
\affiliation{North Carolina A\&T State University, Greensboro, NC 27411, USA}%

\author{D.W.~Higinbotham}%
\affiliation{\JLAB}%

\author{N.D.~Hoffman}%
\affiliation{\CMU}%

\author{P.~Hurck}%
\affiliation{University of Glasgow, Glasgow G12 8QQ, Scotland, United Kingdom}%

\author{I.~Jaegle}%
\affiliation{\JLAB}%

\author{A.~Karki}%
\affiliation{\MSU}%

\author{W.~Li}%
\affiliation{Stony Brook University, Stony Brook, NY 11794, USA}%
\affiliation{CFNS, Stony Brook, NY 11794, USA}%

\author{V.~Lyubovitskij}%
\affiliation{Tomsk State University, 634050 Tomsk, Russia}
\affiliation{Tomsk Polytechnic University, 634050 Tomsk, Russia}%

\author{H.~Marukyan}%
\affiliation{A. I. Alikhanyan National Science Laboratory, Yerevan Physics Institute, Yerevan 0036, Armenia}%

\author{M.D.~McCaughan}%
\affiliation{\JLAB}%

\author{M.E.~McCracken}%
\affiliation{Washington \& Jefferson College, Washington, PA 15301, USA}%

\author{S.~Oresic}%
\affiliation{\UR}%

\author{Z.~Papandreou}%
\affiliation{\UR}%

\author{C.~Paudel}%
\affiliation{\FIU}%

\author{S.~Ratliff}%
\affiliation{\GW}%

\author{E.M.~Seroka}%
\affiliation{\GW}%

\author{S.~Somov}%
\affiliation{National Research Nuclear University MEPhI, Moscow 115409, Russia}%

\author{I.~Strakovsky}%
\affiliation{\GW}%

\author{K.~Suresh}%
\affiliation{\UR}%

\author{A.~Thiel}%
\affiliation{Helmholtz-Institut für Strahlen- und Kernphysik, University of Bonn, Germany}%

\author{B.~Zihlmann}%
\affiliation{\JLAB}%

\date{\today}

\begin{abstract}

We report on the results of the first search for the production of axion-like particles (ALP) via Primakoff production on nuclear targets using the GlueX detector. 
This search uses an integrated luminosity of 100 pb$^{-1}\cdot$nucleon on a $^{12}$C target, and explores the mass region of $200<m_a<450$ MeV via the decay $X\rightarrow\gamma\gamma$. 
This mass range is between the $\pi^0$ and $\eta$ masses, which enables the use of the measured $\eta$ production rate to obtain absolute bounds on the ALP production with reduced sensitivity to experimental luminosity and detection efficiency.
We find no evidence for an ALP, consistent with previous searches in the quoted mass range, and present limits on the coupling on the scale of $\mathcal{O}$(1 TeV).
We further find that the ALP production limit we obtain is hindered by the peaking structure of the non-target-related dominant background in GlueX, which we treat by using data on $^4$He to estimate and subtract these backgrounds.
We comment on how this search can be improved in a future higher-statistics dedicated measurement.
\end{abstract}

\maketitle

\section{Introduction} 

Axion-like particles~(ALPs) are a compelling extension of the standard model~(SM) of particle physics. 
They naturally arise as potential solutions to the strong CP~\cite{Peccei:1977ur,Peccei:1977ur,Weinberg:1977ma,Wilczek:1977pj} and Hierarchy~\cite{Graham:2015cka} problems, and they serve as portal to dark sectors~\cite{Nomura:2008ru,Freytsis:2010ne,Dolan:2014ska,Hochberg:2018rjs}. 
See Refs.~\cite{Marsh:2015xka,Graham:2015ouw,Hook:2018dlk,Irastorza:2018dyq,Choi:2020rgn} for comprehensive reviews.

Since ALPs are pseudo-Nambu-Goldstone bosons, their mass ($m_a$) can be much smaller than the scale $\Lambda$ that controls their interaction with SM particles.  
ALPs at the MeV-to-GeV mass scales have received recent attention~\cite{Bauer:2021mvw,Jerhot:2022chi,Lanfranchi:2020crw,Antel:2023hkf,Agrawal:2021dbo}. 
Such ALPs could predominantly couple to photons, with an effective ALP-photon interaction given by
\begin{align}
    \mathcal{L}_{\rm eff}
    \supset \frac{1}{4\Lambda} a F^{\mu\nu}\tilde{F}_{\mu\nu}\, ,
\end{align}
where $a$ is the axion scalar field and $F^{\mu\nu}$ is the photon field strength tensor with $\tilde{F}^{\mu\nu}=\frac{1}{2}\epsilon^{\mu\nu\alpha\beta}F_{\beta\alpha}$. 
(This assumes a CP-odd pseudoscalar ALP, but the following analysis applies also for a CP-even scalar ALP.)
This interaction with photons serves as a possible portal to probe beyond-SM physics using SM probes and decays.
 
It has been proposed~\cite{Aloni:2019ruo} to search for sub-GeV ALPs with dominant coupling to photons via Primakoff production from nuclei~\cite{PhysRev.81.899}. 
Such a search requires a high-luminosity beam of photons incident on a nuclear target, as well as a large-acceptance detector capable of detecting two final-state photons with a wide range of invariant masses.

The axion and neutral meson ($\pi^0$ and $\eta$) Primakoff differential cross sections are well-known and are similar up to known kinematic function~\cite{Aloni:2019ruo}. 
Therefore, the ALP search can be done in a data-driven manner by normalizing the ALP signal yield to the neutral meson production rate and decay in the diphoton channel.
As a result the dependence on the nuclear form factor and the incident photon beam luminosity cancels, leading to reduced systematic uncertainties.
A previous analysis in GlueX~\cite{Adhikari_2022} has used data on a hydrogen target to search for photoproduced ALP with dominant gluon couplings.

In this work, we report the results of the first exploratory search for ALPs with photon coupling and sub-GeV mass using nuclear targets with the GlueX detector, which recently performed measurements using nuclear targets~\cite{E12:11:003A}.
These data, primarily dedicated to the study of short-range correlations~(SRC) in nuclei~\cite{RevModPhys.89.045002} and color-transparency~(CT)~\cite{Dutta_2013}, studied a number of nuclei, the heaviest of which is $^{12}$C. 
We use these data to realize the ALP search and study the reach of a dedicated future measurement with GlueX.

\begin{figure}[t]
    \centering
    \includegraphics[width = 0.4 \textwidth]{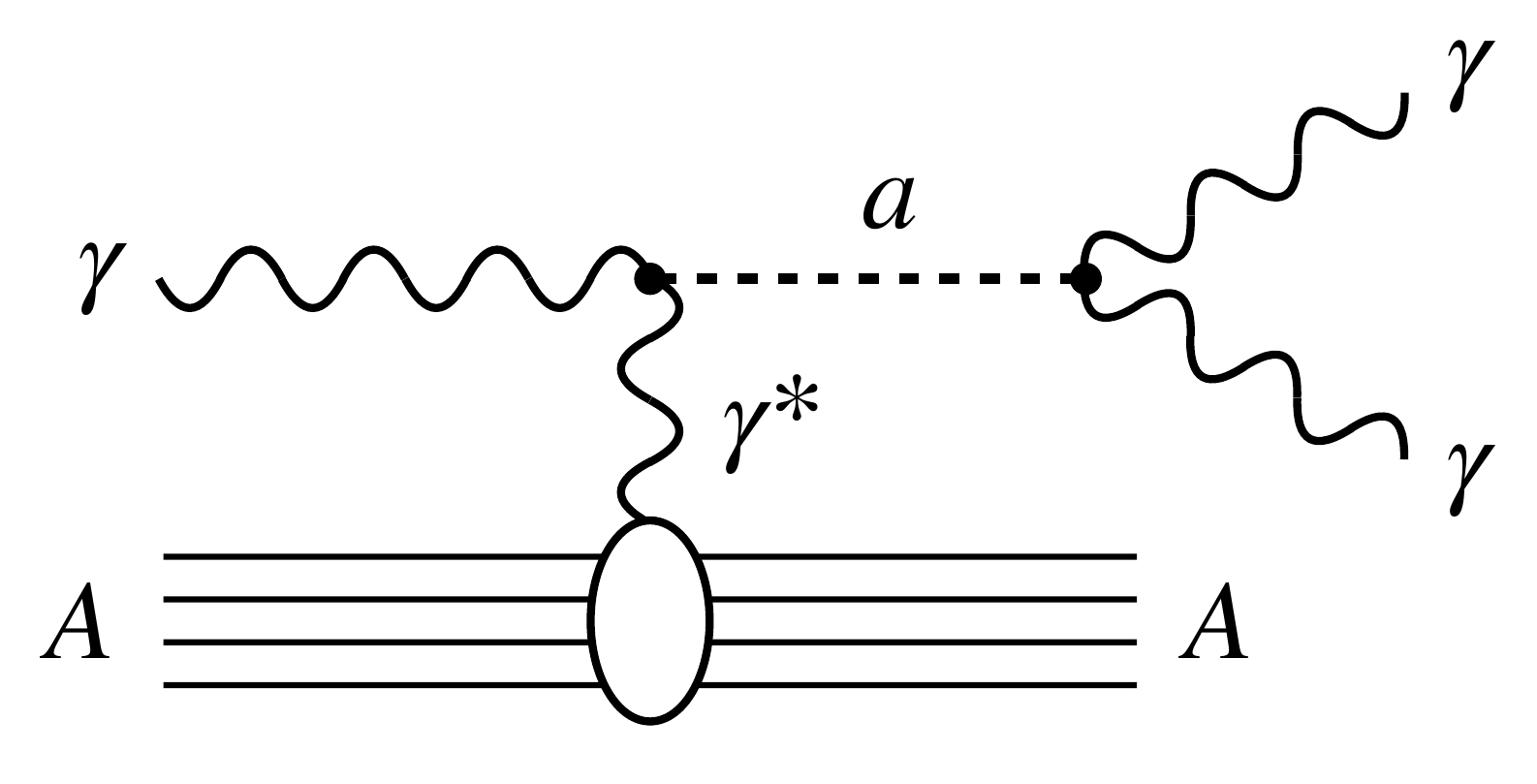}
    \caption{Diagram of the Primakoff production of an ALP, with subsequent decay into $2\gamma$. The incoming beam photon interacts with nucleus $A$ coherently and produces two final-state photons through the mediation of an intermediate spin-0 particle $a$. 
    }
    \label{fig:reaction}
\end{figure}

\section{Experiment} 

The data used in this search were measured using the GlueX spectrometer located in Hall D of the Thomas Jefferson National Accelerator Facility. 
A 10.8\,GeV high-energy electron beam from the Continuous Electron Beam Accelerator Facility~\cite{doi:10.1146/annurev.nucl.51.101701.132327} was used to create a tagged linearly-polarized photon beam via coherent bremsstrahlung from a diamond radiator. 
The energy of the bremsstrahlung photon is deduced from the momentum of the scattered electron measured in the tagging Microscope and Hodoscope detectors~\cite{Hodoscope}. 
This enables a photon-beam energy measurement to an accuracy of about 0.1\%. 
The photon beam is collimated upon exiting the tagger hall, after which it is incident on the target within the GlueX spectrometer. 
In this experiment a solid multifoil $^{12}$C target (8 equidistant foils with a total thickness of 1.9 cm, extended over a 30-cm region) was used, with a total integrated luminosity of $\sim$100 pb$^{-1}\cdot$nucleon.

The GlueX spectrometer~\cite{GlueXNIM} is a large-acceptance detector and includes a number of subdetectors. 
{Immediately surrounding the target is a scintillator-based start counter (SC)~\cite{SC}, a straw-tube central drift chamber (CDC)~\cite{CDC}, a lead and scintillating-fiber barrel calorimeter (BCAL)~\cite{BCAL}, and a superconducting solenoid magnet. 
Further downstream in the direction of the beamline are a set of planar wire forward drift chambers (FDC)~\cite{FDC}, a time-of-flight scintillator detector (TOF), and a lead-glass forward calorimeter (FCAL)~\cite{FCAL}.} 
Physics events in the detector are recorded if sufficient energy is deposited in the calorimeters; a second trigger recorded events with a lower energy threshold in the case of a detected hit in the SC, but was not used in this analysis. 
As the measured final-state consisted solely of two high-energy photons, the calorimeters, specifically the FCAL, provided the majority of the necessary measurements to reconstruct the event, but the other subdetectors were used in the rejection of background processes.

\section{Event Selection} 

This search is based on the Primakoff production of pseudoscalar resonances decaying into 2 photons, $\gamma A\rightarrow AX\rightarrow A\gamma\gamma$. 
In Primakoff production, the 4-momentum transfer $|t|$ to the nucleus is very small, and the mass of the $^{12}$C nucleus means that recoil nuclei cannot be detected. 
As such, the signal events of interest consist of a 2-photon final-state, with no other measured charged or neutral particles.
The photons were measured by observing showers in the FCAL, which reported the energy and the location of the showers. 
Full information of the 4-momentum of the photons $p_{\gamma i}$ was determined by assuming a reaction vertex in the center of the target, neglecting the small target width and allowing us to infer the angle of the photon momentum. 
The total 4-momentum of the 2-photon system $p_X=p_{\gamma1}+p_{\gamma2}$ is further inferred by adding the momentum of the 2-photons, allowing us to calculate the invariant mass and the angle of the ``diphoton'' system.

The event selection criteria, which are enumerated in Table~\ref{tab:cuts}, were first tuned in a blinded analysis of the complete data set. Blinding was achieved by analyzing a 10\% subset of the data.
The specific values used in the background vetoes and the physics cuts were tuned by comparing data to Monte-Carlo simulation of signals in order to optimize the statistical significance of any detection.

\begin{table}[]
\caption{Summary of the event selection criteria. Photon selection criteria were used to select valid decay photon candidates for an event. Vetoes were used to reject background events, and physics cuts were used to select possible Primakoff production events. }
\renewcommand{\arraystretch}{1.1}
\setlength{\tabcolsep}{5pt} 
\begin{tabular}{l|l}
\hline
\multirow{4}{*}{\begin{tabular}[c]{@{}l@{}}
Photon \\ Selection\end{tabular}} & $|t_\text{shower} - t_\text{RF}| < 3$ ns           \\
                                  & $E_\text{shower} > 100$ MeV             \\
                                  & $R_\text{shower} < 105.5$ cm               \\
                                  & Outside Inner FCAL Layer           \\ \hline
\multirow{3}{*}{Vetoes}            & TOF Hit with $|t_\text{tof}-t_\text{shower}|<6.5$ ns \\
                                  & \hspace{1.3cm} and $|{\vec r}_\text{tof}-{\vec r}_\text{shower}|<6$ cm           \\
                                  & Extra FCAL shower with $|t_\text{shower} - t_\text{RF}|<4$ ns \\
                                  & Extra BCAL shower with $|t_\text{shower} - t_\text{RF}|<6$ ns \\ \hline
\multirow{2}{*}{\begin{tabular}[c]{@{}l@{}}Physics \\ Cuts\end{tabular}}     & $0.95 < E_X/E_{\gamma} < 1.05$                         \\
                                  & $\theta_X < 0.5^\circ$      \\ \hline
\end{tabular}\label{tab:cuts}
\end{table}

Events were required to have exactly two neutral shower candidates satisfying four criteria. 
First, the showers needed to originate from within 3\,ns of the electron-beam RF time for the event, accounting for the expected time-of-flight. 
Second, the showers were required to have a measured energy greater than 100\,MeV. 
Third, the showers were required to be located outside the innermost layer of the FCAL closest to the beamline.
Finally, the showers were required to be within 105.5\,cm of the center of the FCAL.
The events were also required to have at least one tagged beam photon candidate within 2\,ns of the RF time, after accounting for time-of-flight to the target.

Several veto conditions were checked to remove possible background events. 
Events with a hit in the TOF scintillator in proximity to the calorimeter shower were rejected to remove charged-particle backgrounds. 
Events with additional showers in either the FCAL or BCAL were rejected in order to suppress non-Primakoff events with additional particles.

Several physics cuts were applied to the events to isolate Primakoff contributions. 
An ``elasticity'' cut was applied, requiring that the total energy of the two detected photons be within 5\% of the beam photon energy ($0.95<E_X/E_\gamma<1.05$), in order to reduce inelastic contributions and limit accidental photon contribution.
An additional cut was placed on the angle $\theta_X$ of the diphoton relative to the beamline; the 2-photon system was required to have a small angular deflection $\theta_X<0.5^\circ$, restricting the data to a region where Primakoff contributions dominate.
These latter two cuts had the combined impact of limiting the missing mass of the reaction to be close to the mass of an intact carbon nucleus, though the resolution of the detector did not allow precise enough reconstruction to cut on the missing mass.

\begin{figure}[t]
    \centering
    \includegraphics[width = 0.45 \textwidth]{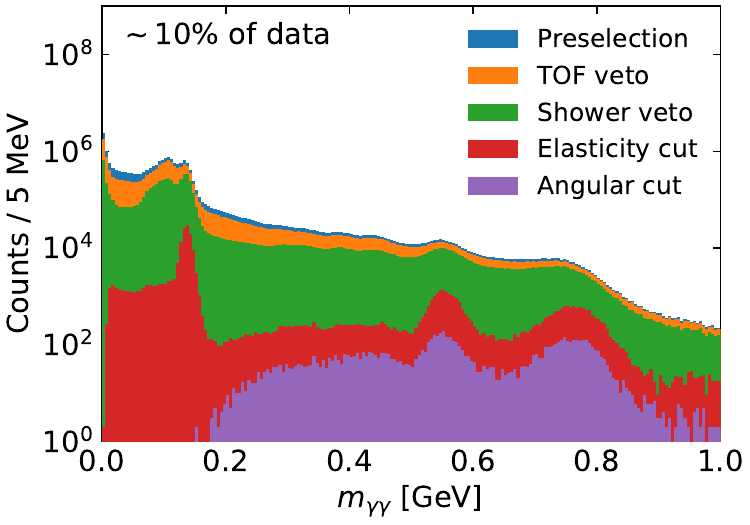}
    \caption{The invariant 2-photon mass spectrum in the blinded subset of data at each level of cut and veto, applied sequentially.}
    \label{fig:impact}
\end{figure}

Fig.~\ref{fig:impact} shows the effect of selection vetoes and cuts on the invariant 2-photon mass spectrum. 
We note that the $\eta$ meson peak at the 2-photon invariant mass $m_{\gamma\gamma}=548$\,MeV is clearly seen after all selection criteria have been applied, allowing the search to be normalized relative to this channel. 
However, the $\pi^0\rightarrow\gamma\gamma$ events are ultimately removed upon application of the angular deflection cut. 
This is an acceptance effect; the mass of the diphoton correlates with the opening angle of the photons, and requiring the diphoton system to have a small deflection angle means that photons originating from low-mass diphotons do not impact sufficiently far from the beamline to fall within the FCAL. 
This results in a sharp loss of signal below an invariant mass of $m_{\gamma\gamma}\approx180$\,MeV.

We also observe a peak above the $\eta$ meson in mass, which corresponds to the decay $\omega\rightarrow\gamma\pi^0\rightarrow\gamma(\gamma\gamma)$. 
In a sizeable fraction of events, at least one photon from this decay is undetected, creating the appearance of a 2-photon final-state. This large background limits searches for resonances in the region $m_{\gamma\gamma}>m_\eta$.

\section{Statistical Analysis} 

We performed a bump-hunt on the 2-photon mass spectrum in the diphoton invariant mass range of 200 to 450 MeV. 
This lower bound is near the limit of detector acceptance for Primakoff events, and the upper bound is proximate to the $\eta$ peak.

The distribution of 2-photon resonance signal is seen in simulation to follow a Gaussian shape, and the resolution of this Gaussian $\sigma_m(m_a)$ was taken from simulation for a given $m_a$ hypothesis; 
in general, the mass resolution in the search range was found to be $3-4\%$ and to be roughly constant with $m_a$. 
The simulated mass resolution was found to agree with that measured for the decay $\eta\rightarrow\gamma\gamma$.
The background 2-photon combinations was modelled by a polynomial of 4$^{th}$ order

\begin{figure}[t]
    \centering
    \includegraphics[width = 0.45 \textwidth]{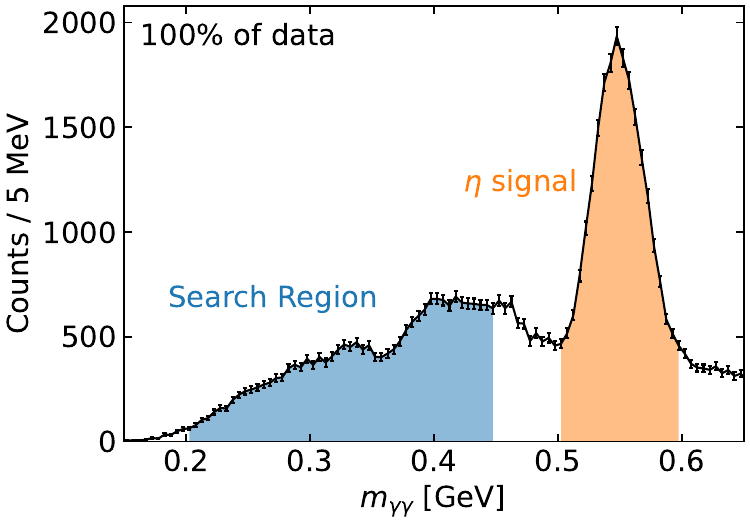}
    \caption{The invariant 2-photon mass spectrum used in the bump hunt after all selection cuts have been applied, including the full set of carbon data. The search region between 200 and 450 MeV is shaded in blue, and the $\eta\rightarrow\gamma\gamma$ signal used for normalization is shaded in orange.}
    \label{fig:shaded}
\end{figure}

For a given mass hypothesis $m_a$, the measured 2-photon mass spectrum was considered in a window of width $\Delta m=20\sigma_m$, where $\sigma_m$ is the 2-photon mass resolution at the test mass. 
This window was centered on $m_a$ when possible, but was not allowed to extend above a value of 500\,MeV to avoid the $\eta$ peak, the modeling of which would otherwise dominate the goodness-of-fit. 
A similar lower bound on the fit region was placed at 180\,MeV due to lack of acceptance below that. 
The data within the search region was filled into 400 bins, giving a bin width of $\sigma_m/20$.
For each $m_a$, the Gaussian signal and polynomial background were fit to data using a maximum likelihood fit. 
The total yield $\mu$ of the signal was allowed to vary, with the mass and width from simulation remaining fixed. The polynomial coefficients of the background were also fit, giving a total of six free parameters.
This model was used to calculate the limits on the signal yield following the procedures of Ref.~\cite{Cowan2013}.
The background-only hypothesis, with $\mu=0$ enforced, was also used to predict the expected limits from the data and their level of fluctuation.

\section{Normalization}
\label{sec:norm}

The yield $\mu_a$ may be related to the ALP-photon $1/\Lambda$ coupling by normalizing relative to the $\eta\rightarrow\gamma\gamma$ yield. 
We note that the signal yield $\mu$ for either process can be expressed as
\begin{align}
   \mu_X=\mathcal L\times \sigma_X\times\epsilon\times\mathcal B(X\rightarrow \gamma\gamma) \,.
\end{align}
Here $\mathcal L$ is the total integrated luminosity, $\sigma_X$ is the total Primakoff production cross section for $X=\pi^0$, $\eta$, $a$. $\epsilon=N_{\rm detected}/N_{\rm total}$ is the total detection and selection efficiency, which depends on mass of the produced pseudoscalar, and $\mathcal B(X\rightarrow \gamma\gamma)$ is the branching ratio of decay into 2-photons, assumed to be 100\% for the ALP and measured to be $39.36\pm0.18$\% for the $\eta$~\cite{PDG}.

By equating the luminosity for the cases of $X=a,\eta$, we derive the relationship between the ALP exclusion and measurement of Primakoff $\eta$: 

\begin{align}\label{eq:rat}
    \sigma_a=\frac{\epsilon_\eta \mu_a}{\epsilon_a \mu_\eta}\mathcal B(\eta\rightarrow \gamma\gamma) \times \sigma_\eta \, .
\end{align}

The Primakoff cross section can be factorized into a nuclear form factor, the photon coupling $1/\Lambda$ and a purely kinematic component which depends on the resonance mass (see Ref.~\cite{Aloni:2019ruo}), i.e. $\sigma_X=\frac{1}{\Lambda_X^2} \sigma_0(m_X)$.
By encompassing the mass-dependent cross section and efficiency effects into a single factor, 
we may relate the excluded ALP-photon coupling to the $\eta$-photon coupling, which can be calculated from the measured $\eta\rightarrow \gamma\gamma$ partial width $\Gamma_{\eta\rightarrow\gamma\gamma}=m_\eta^3/(64\pi\Lambda^2_\eta)=520\pm20$ eV~\cite{PDG}:

\begin{align}
    \frac{1}{\Lambda_{95}}
 =   8{\mathcal B}(\eta\rightarrow \gamma\gamma) \sqrt{\frac{\sigma_0(m_\eta)}{\sigma_0(m_a)}\frac{\epsilon(m_\eta)}{\epsilon(m_a)}\frac{\mu_{a,95}}{\mu_\eta}  \frac{\pi \Gamma_{\eta}}{m_\eta^3}}
\,,
    \label{eq:coupling}
\end{align}
where $\mu_{a,95}$ and $1/\Lambda_{95}$ are the 95\,\% upper bounds on the ALP yield and on the ALP photon coupling, respectively, and $\Gamma_\eta$ is the total $\eta$ decay width.

One must also consider that normalization to the $\eta$ meson yield must be performed specifically relative to the number of \emph{Primakoff} $\eta\rightarrow \gamma\gamma$ events.
In contrast, the measurement of $\eta\rightarrow\gamma\gamma$ also includes contributions from incoherent nuclear production, coherent nuclear production, and interference between coherent and Primakoff production.
The restriction to a diphoton scattering angle of $\theta_X<0.5^\circ$ serves to reduce contributions from these other production mechanisms, which are more dominant at larger scattering angles, but does not entirely eliminate them. 
An overestimate of the Primakoff $\eta$ meson yield, as one can see from Eq.~\ref{eq:coupling}, would result in an overly aggressive claim of the upper limit set by the data.

\begin{figure}[t]
    \centering
    \includegraphics[width = 0.45 \textwidth]{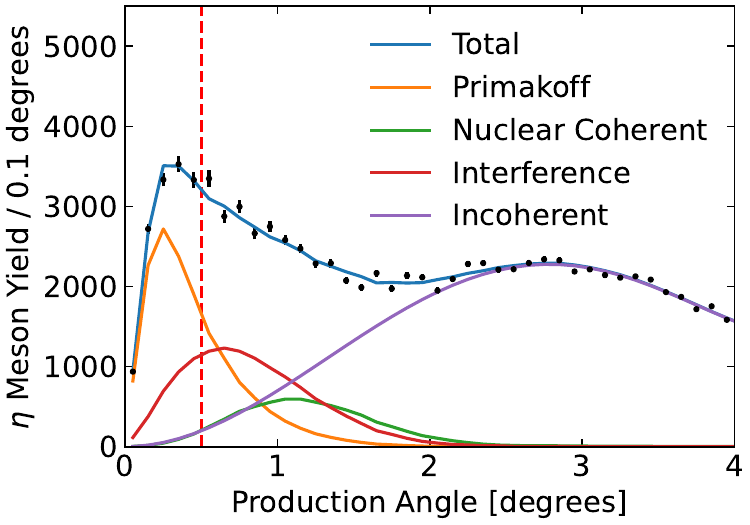}
    \caption{The extracted yield of $\eta\rightarrow\gamma\gamma$ events in different bins of the angular deflection of the $\eta$. 
    We observe a distinct Primakoff peak at $\theta_X<0.5^\circ$ as well as a substantial contribution of incoherent events dominating at $\theta_X\sim3^\circ$.
    The angular distribution is fit to a sum of the different contributions to $\eta$ photoproduction, including Primakoff, nuclear coherent, interference between the two, and nuclear incoherent.
    }
    \label{fig:eta_theta}
\end{figure}

To estimate the yield of $\eta\rightarrow\gamma\gamma$ events resulting from Primakoff production, we examine the angular distribution of these events, shown in Fig.~\ref{fig:eta_theta}. 
These event yields are obtained by fitting the mass spectrum for each angular bin in the region $450<m_{\gamma\gamma}<650$ MeV using a Gaussian signal with a linear background, which is found to perform well at larger deflection angles, and relaxing only the angular cut on the data.
We observe a sharp peak in the $\eta\rightarrow\gamma\gamma$ yield at $\theta_X<0.5^\circ$, which corresponds to Primakoff production, but we also see substantial contributions of events at larger angles. 
In particular, a significant contribution of events come from a wider distribution centered at $\theta_X\sim3^\circ$, corresponding to nuclear incoherent production of $\eta$ mesons. 

The fraction of $\eta$ resulting from Primakoff production was estimated by fitting this angular distribution with the contributions of the four production mechanisms. 
These include Primakoff production, nuclear coherent production (modeled using the calculations of Ref.~\cite{PhysRevC.80.055201}), the interference between the two, and incoherent photoproduction. 
The contribution from incoherent photoproduction was modelled using a 5$^{th}$-order polynomial fit, with the value and slope at $\theta_X=0$ both constrained to be zero.

This fit results in an estimate of $72\pm6_{\rm stat}\pm10_{\rm sys}$\ \% contribution from Primakoff production in the region of interest $\theta_X<0.5^\circ$. 
We assign the $10\%$ systematic uncertainty to account for model uncertainties, particularly in the description of the incoherent production; this was assessed by testing different models and constraints for the description of the incoherent component. 
These uncertainties are listed in Table~\ref{tab:norm}, along with systematic uncertainties relating to the decay of the $\eta$.
The total normalization uncertainty on the excluded ALP cross section is found to be 17\%.
Systematic uncertainties from the efficiency of detecting the diphoton are assumed to be considerably smaller than the dominant uncertainties from $\eta$ production~\cite{Adhikari_2022}, and are further reduced by taking ratio between two similar final-states as in Eq.~\ref{eq:rat}; as such, this uncertainty is neglected in the calculation of the limits.

\begin{table}[]
\caption{Summary of the normalization uncertainties impacting the excluded ALP cross section. These uncertainties are dominated by those relating to the extraction of the Primakoff $\eta\rightarrow\gamma\gamma$ yield as described in the text. Also included are uncertainties on the $\eta$ total width and branching ratio to $\gamma\gamma$, taken from Ref.~\cite{PDG}.}
\begin{tabular}{l|l}
Source                                                                         & Uncertainty \\ \hline
Primakoff $\eta$ yield (statistical)                                           & 8\%         \\
Primakoff $\eta$ yield (systematic)                                            & 14\%        \\
$\Gamma_{\eta}$                                                                & 4\%         \\
$\mathcal{B}(\eta\rightarrow\gamma\gamma)$                                     & 0.7\%       \\ \hline
Total                                                                          & 17\%       
\end{tabular}\label{tab:norm}
\end{table}

\section{Backgrounds}

\begin{figure}[t]
    \centering
    \includegraphics[width = 0.5 \textwidth]{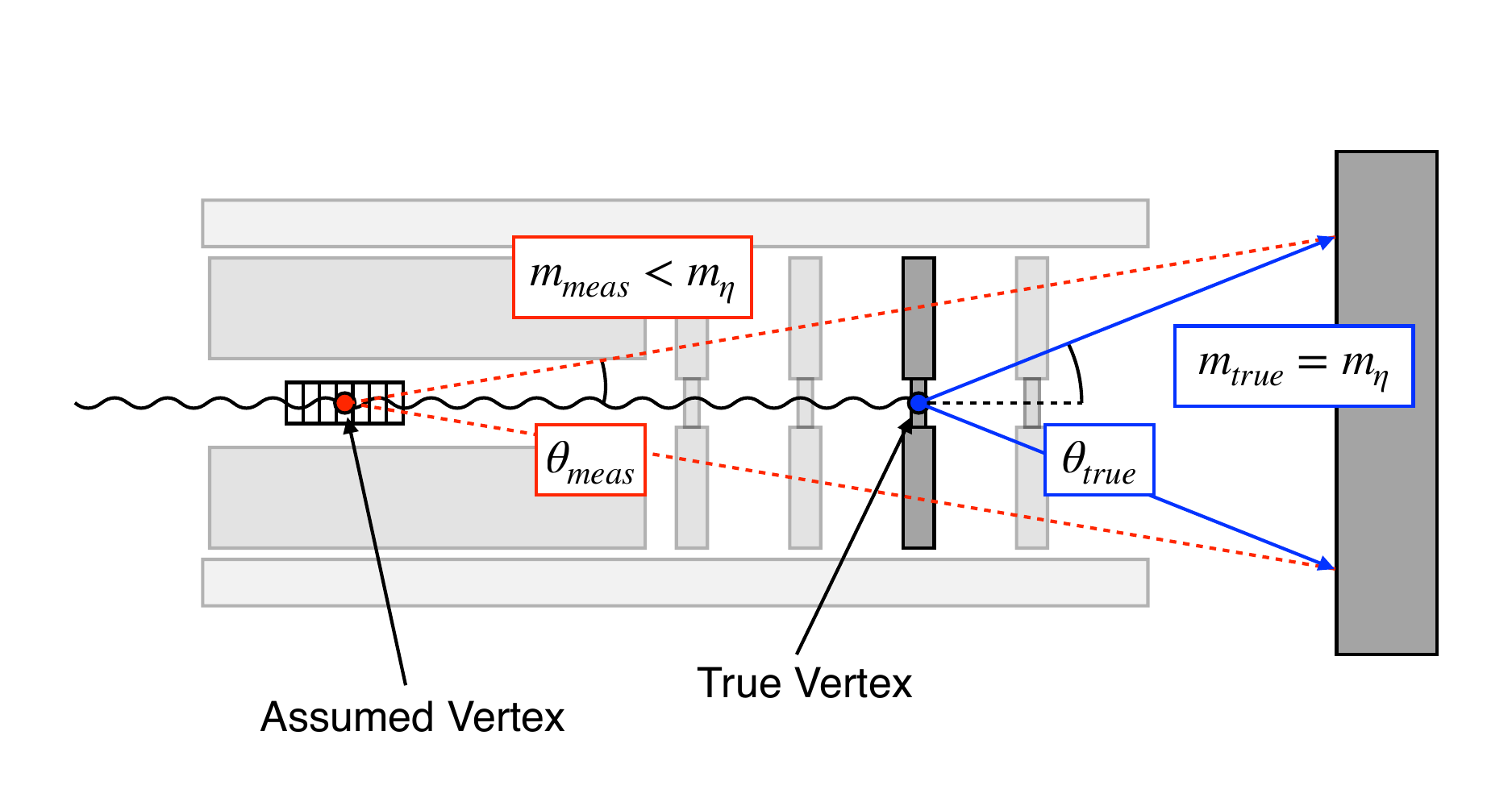}
    \caption{Example of a background event resulting from downstream in-beamline material. An $\eta$ meson is produced in the FDC package material and decays into two photons, which impact the FCAL at a given opening angle $\theta_{\rm true}$ correlating to their invariant mass. The energy and location of each photon shower is measured in the FCAL, but the assumed vertex within the target results in an underestimated opening angle $\theta_{meas}<\theta_{true}$. The reconstructed 2-photon mass is thus below that of the true $\eta$ mass. Similar background events can occur from other FDC packages or the air downstream of the target. $\omega\rightarrow \pi^0\gamma$ production is also possible.}
    \label{fig:BG_diagram}
\end{figure}

It is important to explore the primary source of background for this measurement, which is the photoproduction of $\eta\rightarrow \gamma\gamma$ and $\omega\rightarrow\pi^0\gamma$ outside of the target. 
Fig.~\ref{fig:BG_diagram} shows an example of such a background event. 
In this event, the $\eta$ meson is photoproduced in material downstream of the target from the beam photon, and decays into two photons. 
These photons impact the FCAL, with their energy deposition and shower locations being measured. 
The interaction vertex, however, cannot be isolated due to the lack of charged tracks in the event, and must be assumed to take place in the center of the target. 
This misplaced vertex results in an underestimated opening angle between the photons, and therefore in an underestimation of the invariant 2-photon mass as well. 
Events of this type, including both $\eta\rightarrow\gamma\gamma$ and misreconstructed $\omega\rightarrow\pi^0\gamma$ events, can result in reconstructed invariant masses in the search region of $200\,{\rm MeV} < m_{X} < 450$\,MeV.

In the event that downstream material within the beamline is completely evenly distributed, such as in the case of air within the experimental hall, these processes would result in substantial but smoothly varying backgrounds, reducing the sensitivity of the search but not requiring detailed background modeling.
However, not all excess material in the experimental hall is evenly distributed, and the most concerning backgrounds come from the FDC. 
Each of the 4 FDC packages has about 0.22\% radiation lengths of material directly within the photon beamline. 
This material is less than the total amount of air in the chamber, which is on the order of 1.8\% radiation lengths in the region between the target and the FCAL, but is of far greater concern due to its concentration at a specific point in the spectrometer. 
Background processes from the FDC result not in smoothly distributed background, but sharp features in the mass spectrum corresponding to the locations of the FDC in the hall.

These sharp features could result in large deviations in the coupling limit set by the assumption of polynomial background.
Any complex background structure could result in both false discovery of apparent resonances and in overestimates of the coupling limits.
While with it is possible to address the background by theoretical modeling, such models require detailed understanding of the different $\eta$ and $\omega$ production mechanisms at these energies as well as understanding the distribution of material in the beamline, and would involve a large number of parameters to be fit to data.
This would introduce considerable model-dependency in the extraction, and is further complicated by the peaking nature of this background, which causes significant degeneracy between background and signal shapes for a large fraction of the mass range.
For the purposes of this analysis, such modeling was not performed, and the background was fit using the previously described polynomial function.

In order to treat the effects of beamline-material background, data without the carbon target present can be used to subtract the mass spectrum and remove contributions in common. 
While little empty-target data was measured for this experiment, substantial data was taken with other nuclear targets.
In particular, a comparable fraction of data, roughly 67 pb$^{-1}\cdot$nucleon, was taken on a $^4$He target. 
As the per-nucleon Primakoff cross section for helium differs from that for carbon by a calculated factor of $\frac{\sigma_{He}/4}{\sigma_{C}/12}=0.38$, the 2-photon mass spectrum for this data can be subtracted from the spectrum for carbon to account for common non-target-related backgrounds without substantially reducing the sensitivity to Primakoff production of ALP.
The helium data is scaled by the ratio of the total number $f$ of photons on each target, measured by the Pair Spectrometer~\cite{BARBOSA2015376} to be $f_C/f_{He}=1.65$; as experimental effects are common between each set of data, this ratio is understood to much higher precision than the individual photon flux per target.

\begin{figure}[th]
    \centering
    \includegraphics[width = 0.45 \textwidth]{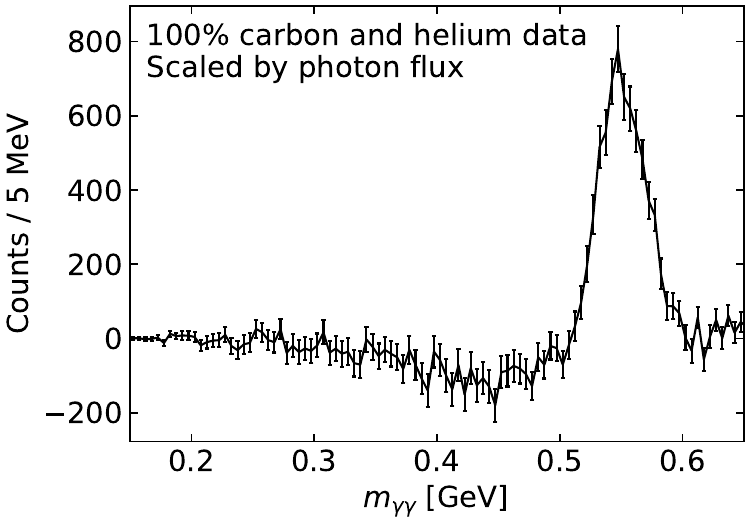}
    \caption{The invariant 2-photon mass spectrum for all carbon data after subtraction of helium data. Helium data was scaled by the ratio of the total photon flux for the data on each target. Subtracted data in the search region $200<m_{X}<450$ MeV is found to be very flat and smooth, supporting the hypothesis that primary background are the result of in-beam material which is common between the targets.}
    \label{fig:subtracted}
\end{figure}

Fig.~\ref{fig:subtracted} shows the 2-photon mass spectrum for the full carbon data after subtracting the scaled helium data. 
We observe that mass spectrum is considerably smoother in the search region following the subtraction, with the complex background structures no longer present. 
This observation supports the hypothesis that the primary backgrounds are beamline-related and therefore common between the targets.
This subtracted spectrum may also be used to perform a bump-hunt following the procedures of Refs.~\cite{Bressler_2014,Aad_2017} where the level of background in each bin is considered a separate nuisance parameter.

We note that, as Primakoff ALP production is suppressed in helium relative to carbon but not negligible, the sensitivity of the subtracted spectrum to ALP signal is reduced by a factor
\begin{equation}
    1 - \frac{f_C}{f_{He}}\frac{L_{He}\sigma_{He}}{L_{C}\sigma_{C}}=0.59\pm0.02
\end{equation}
where $L_{A}$ is the luminosity and $\sigma_{A}$ the Primakoff ALP cross section for nucleus $A$.
Any extracted yields or limits for the ALP signal from the subtracted spectrum are divided by this factor to account for the loss in sensitivity.

\section{Results}

\begin{figure}[t]
    \centering
    \includegraphics[width = 0.45 \textwidth]{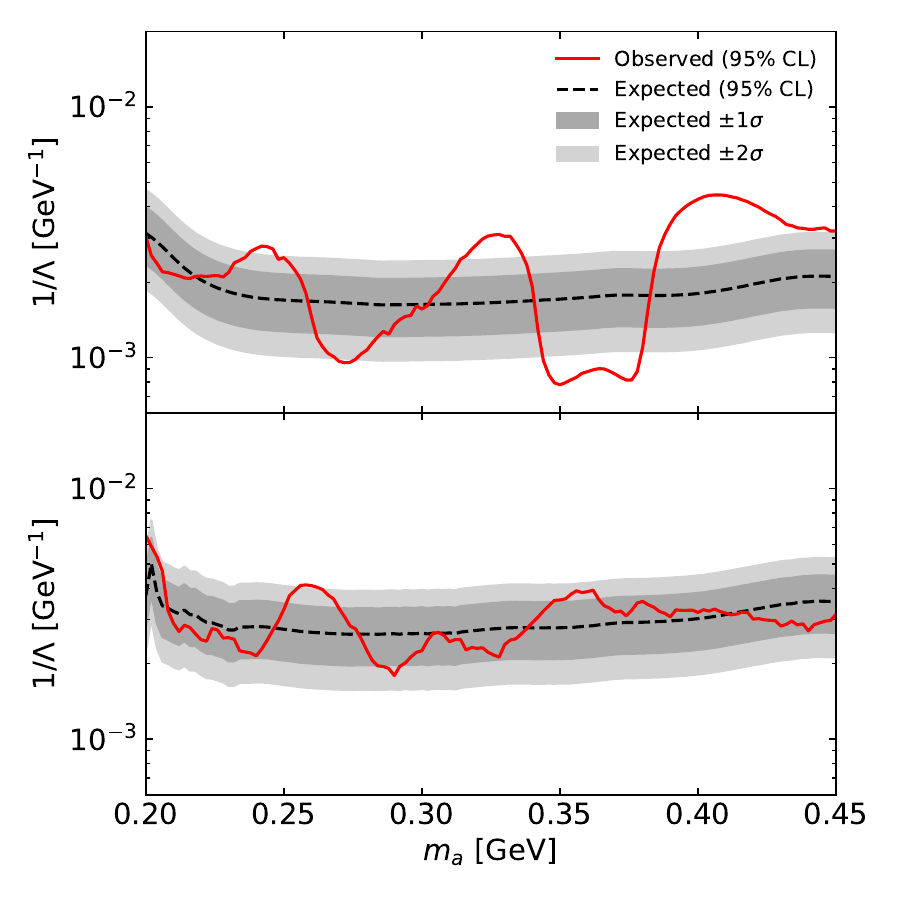}
    \caption{\emph{Top}: The limits calculated by the bump-hunting procedure on the full mass spectrum are compared with those predicted from the background-only hypothesis. The observed limits (solid red line) are shown to be at the scale of those predicted by the background-only hypothesis (dashed line and shaded regions), but fluctuate more strongly than expected, perhaps due to more complex resolved background structure. The most stringent apparent limit is at 360 MeV, resulting from a corresponding dip in the mass spectrum.
    \emph{Bottom}: Same as top, but instead using the background-subtracted mass spectrum. These observed limits agree well with expectations, fluctuating at a level consistent with predictions.}
    \label{fig:coupling}
\end{figure}

The upper limits on the ALP-photon coupling are extracted from the full dataset using the statistical method and normalization to the $\eta\rightarrow\gamma\gamma$ previously described. 
Fig.~\ref{fig:coupling} (top) compares this nominal extracted upper limit with that projected from the background-only fit to the full carbon data, as well as the predicted level of fluctuation in this limit.
We observe that the extracted limits using this procedure agree with the scale predicted by the background-only hypothesis, but can fluctuate beyond the level expected from purely statistical variation.

We see that the most stringent apparent limit is set at an ALP mass of $\sim$360\,MeV, representing a downward fluctuation as compared with the expected limits.
This indicates that the full data may resolve features of the background which cannot be well-described by a simple polynomial fit. 
In the particular case of the 360\,MeV hypothesis, we note that the mass spectrum has a significant dip at this location, which results in a strict apparent limit. This could otherwise indicate that the model used for describing the background requires greater complexity to set accurate limits.

Fig.~\ref{fig:coupling} (bottom) shows the same comparison between the extracted and predicted coupling limits using the background-subtracted mass spectrum.
We find that both the scale and the level of fluctuation in the observed limit agree with the background-only predictions, indicating that the background-subtraction leaves a smooth and well-behaved mass spectrum.
We note that the scale of the limits set by the subtracted spectrum are above those for the unsubtracted spectrum, as the background-subtraction introduces greater uncertainty on the spectrum and also reduces the ALP sensitivity by subtracting any contributions from the helium data.

\begin{figure}[t]
    \centering
    \includegraphics[width = 0.45 \textwidth]{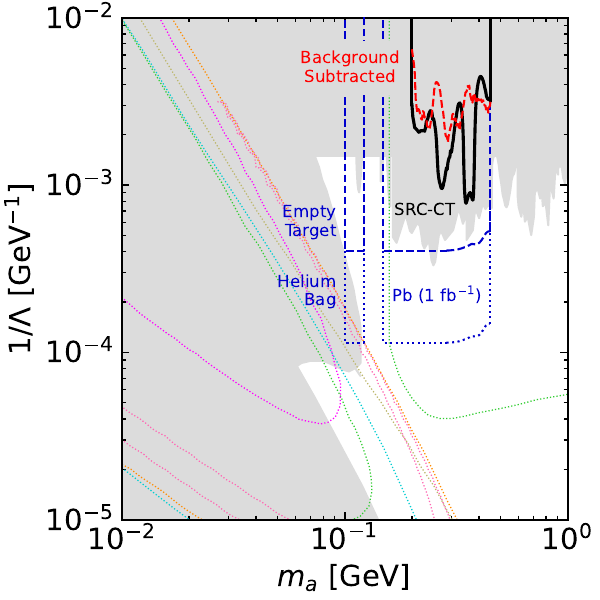}
    \caption{The limits set by this study using the full (solid black) and background-subtracted (dashed red) data are shown alongside the projections for 1 fb$^{-1}\cdot$nucleon of luminosity using a lead target for both the cases of using empty-target subtraction (dashed) and helium balloon placement downstream (dotted).
    For the latter a region around the $\pi^0$ mass is excluded as proximity to this resonance would complicate a bump hunt.
    These are compared with existing limits on ALP coupling as a function of mass (grey shaded region)~\cite{Aloni:2019ruo,OPAL,PhysRevLett.118.171801,PhysRevD.38.3375,BeamDumps,BESIII:2022rzz,mitridate2023energetic} and predicted limits for NA62 (cyan), SHip (dark yellow)~\cite{SHiP}, FASER (magenta)~\cite{PhysRevD.98.055021}, SeaQuest (orange)~\cite{PhysRevD.98.035011}, Belle-II (green)~\cite{BelleII}, and LUXE (pink)~\cite{Bai_2022,Abramowicz_2021}). 
    The results of this study are surpassed by current world-leading limits, while the projections for a lead target and improved acceptance are found to surpass current limits and reach untested regions of parameter space.
    }
    \label{fig:limits}
\end{figure}

Fig.~\ref{fig:limits} shows these extracted limits from the full (solid black) and background-subtracted (dashed red) data compared with current world-leading limits on the parameter space (shaded grey), as well as the expected limits for other experiments. 
We find that the limits set on the coupling by these data are on the scale of $\mathcal{O}$(1 TeV), competitive with recent results.
However, these limits are surpassed by the most recent world-leading limits from BESIII~\cite{BESIII:2022rzz}, which cover a similar range of ALP masses and reach to weaker couplings.

A dedicated Primakoff ALP search at GlueX would require a means of accounting for the off-vertex background. 

Given the challenges in accurately modeling this background, it would be ideal to address it using experimental solutions.
One possible solution is to measure a substantial amount of data without a target present, allowing for precise measurement of the non-target-related backgrounds.
By measuring these to a high precision it would be possible to subtract out the impact of downstream material. 
Such ``empty-target'' data would require a comparable luminosity to the measurement itself to avoid dominating the statistical uncertainties, but the reduced material could allow running at higher photon flux.

A more complete solution would be the removal of the FDC packages from the spectrometer for the duration of the run, and the placement of a helium balloon between the target and the FCAL. 
The helium would present fewer radiation lengths than air by a factor of 40, and the removal of the FDC material would result in a much smoother background profile.
This solution would allow for much greater sensitivity, as the statistical fluctuations in the background would have considerably reduced impact on the sensitivity.

The ALP mass range would naturally be extended to include much lower mases in future experiments due to an upgrade of the FCAL currently in progress~\cite{JEF}. 
The central part of the FCAL is being replaced with an insert, constructed from lead tungstate, which will provide increased energy and position resolution and lower the minimum detection angle to $\sim 0.45^\circ$, considerably lower than photon detection possible in this analysis.
A future experiment, using this calorimeter, would have improved sensitivity due to improved resolution and the search could be extended down to ALP masses as low as $100$ MeV.

We also use our measured data to perform an estimate for a high-luminosity (1 fb$^{-1}\cdot$nucleon) measurement of a lead target, which would be the optimal possibility for performing this measurement in GlueX. 
Using the background-only projections for the limits and scaling the limits appropriately by the ratio of the per-nucleon Primakoff cross section for the two nuclei, $\frac{\sigma_{Pb}/208}{\sigma_{C}/12}\approx7.25$
(calculated using the known Primakoff cross section and integrating over the tagged photon flux), luminosity, and level of background, we calculate the projected limits also shown in Figure~\ref{fig:limits}, including both the case where empty-target data has been collected at comparable statistics to the target data, and the case where the FDC have been removed and replaced with a helium bag.
In both cases we have extended the mass range to include the acceptance of the FCAL insert for 2-photon events, assuming a level of background similar to that measured in these data.
We find that a measurement using empty-target subtraction would provide limits comparable to the current BESIII~\cite{BESIII:2022rzz} limits in the same mass region, and would extend to include the mass region $100<m_a<150$\,MeV, which is otherwise difficult to measure outside the use of beam dumps.
Performing the same measurement after removing the FDC and implementing a helium bag would result in considerably improved sensitivity across the mass range, and would allow exploring certain regions of parameter space covered by neither the Belle-II nor beam dump measurements.
We note that these projected limits are less optimistic than those presented in Ref.~\cite{Aloni:2019ruo}.
This comes about from a combination of more detailed handling of the potential backgrounds as well as more precise estimates of the feasible luminosity for such a measurement.

\section{Conclusions}

In summary, we present a proof-of-principle analysis of the proposed ALP search of Ref.~\cite{Aloni:2019ruo}, using high-energy photon-nucleus data to examine ALP hypotheses within the mass range between $200$ and $450$ MeV. 
We successfully extract limits using current data from a $^{12}$C target, and find that the obtained limits are less stringent than recent world-leading extractions in the studied mass range. 
We identify a number of experimental challenges that currently limit the ability of the GlueX detector to perform such a search, particularly related to the substantial material in the detector which intersects the beamline. 
We demonstrate how the use of multiple targets or empty-target data can be used to mitigate these background and provide more stable exclusion limits.
We provide estimates of the limits that could be set using a longer run with a lead target and an improved experimental setup, which could provide world-leading limits over a range of possible ALP masses.

\begin{acknowledgments}

We acknowledge the efforts of the staff of the Accelerator and Physics Divisions at Jefferson Lab that made this experiment possible.
We thank M. Williams for many valuable discussions.
This work was supported in part by the U.S. Department of Energy, the U.S-Israel Binational Science Foundation, the Israeli Science Foundation, the Pazy Foundation, and the Azrieli Foundation.
Jefferson Science Associates operates the Thomas Jefferson National Accelerator Facility for the DOE, Office of Science, Office of Nuclear Physics under contract DE-AC05-06OR23177.
This research also used resources of the National Energy Research Scientific Computing Center (NERSC), a U.S. Department of Energy Office of Science User Facility operated under Contract No. DE-AC02-05CH11231.

\end{acknowledgments}

\bibliography{references}

\begin{thebibliography}{51}%
\makeatletter
\providecommand \@ifxundefined [1]{%
 \@ifx{#1\undefined}
}%
\providecommand \@ifnum [1]{%
 \ifnum #1\expandafter \@firstoftwo
 \else \expandafter \@secondoftwo
 \fi
}%
\providecommand \@ifx [1]{%
 \ifx #1\expandafter \@firstoftwo
 \else \expandafter \@secondoftwo
 \fi
}%
\providecommand \natexlab [1]{#1}%
\providecommand \enquote  [1]{``#1''}%
\providecommand \bibnamefont  [1]{#1}%
\providecommand \bibfnamefont [1]{#1}%
\providecommand \citenamefont [1]{#1}%
\providecommand \href@noop [0]{\@secondoftwo}%
\providecommand \href [0]{\begingroup \@sanitize@url \@href}%
\providecommand \@href[1]{\@@startlink{#1}\@@href}%
\providecommand \@@href[1]{\endgroup#1\@@endlink}%
\providecommand \@sanitize@url [0]{\catcode `\\12\catcode `\$12\catcode
  `\&12\catcode `\#12\catcode `\^12\catcode `\_12\catcode `\%12\relax}%
\providecommand \@@startlink[1]{}%
\providecommand \@@endlink[0]{}%
\providecommand \url  [0]{\begingroup\@sanitize@url \@url }%
\providecommand \@url [1]{\endgroup\@href {#1}{\urlprefix }}%
\providecommand \urlprefix  [0]{URL }%
\providecommand \Eprint [0]{\href }%
\providecommand \doibase [0]{https://doi.org/}%
\providecommand \selectlanguage [0]{\@gobble}%
\providecommand \bibinfo  [0]{\@secondoftwo}%
\providecommand \bibfield  [0]{\@secondoftwo}%
\providecommand \translation [1]{[#1]}%
\providecommand \BibitemOpen [0]{}%
\providecommand \bibitemStop [0]{}%
\providecommand \bibitemNoStop [0]{.\EOS\space}%
\providecommand \EOS [0]{\spacefactor3000\relax}%
\providecommand \BibitemShut  [1]{\csname bibitem#1\endcsname}%
\let\auto@bib@innerbib\@empty
\bibitem [{\citenamefont {Peccei}\ and\ \citenamefont
  {Quinn}(1977)}]{Peccei:1977ur}%
  \BibitemOpen
  \bibfield  {author} {\bibinfo {author} {\bibfnamefont {R.~D.}\ \bibnamefont
  {Peccei}}\ and\ \bibinfo {author} {\bibfnamefont {H.~R.}\ \bibnamefont
  {Quinn}},\ }\bibfield  {title} {\bibinfo {title} {"{Constraints Imposed by CP
  Conservation in the Presence of Instantons}"},\ }\href@noop {} {\bibfield
  {journal} {\bibinfo  {journal} {Phys. Rev. D}\ }\textbf {\bibinfo {volume}
  {16}},\ \bibinfo {pages} {1791} (\bibinfo {year} {1977})}\BibitemShut
  {NoStop}%
\bibitem [{\citenamefont {Weinberg}(1978)}]{Weinberg:1977ma}%
  \BibitemOpen
  \bibfield  {author} {\bibinfo {author} {\bibfnamefont {S.}~\bibnamefont
  {Weinberg}},\ }\bibfield  {title} {\bibinfo {title} {"{A New Light
  Boson?}"},\ }\href@noop {} {\bibfield  {journal} {\bibinfo  {journal} {Phys.
  Rev. Lett.}\ }\textbf {\bibinfo {volume} {40}},\ \bibinfo {pages} {223}
  (\bibinfo {year} {1978})}\BibitemShut {NoStop}%
\bibitem [{\citenamefont {Wilczek}(1978)}]{Wilczek:1977pj}%
  \BibitemOpen
  \bibfield  {author} {\bibinfo {author} {\bibfnamefont {F.}~\bibnamefont
  {Wilczek}},\ }\bibfield  {title} {\bibinfo {title} {"{Problem of Strong $P$
  and $T$ Invariance in the Presence of Instantons}"},\ }\href@noop {}
  {\bibfield  {journal} {\bibinfo  {journal} {Phys. Rev. Lett.}\ }\textbf
  {\bibinfo {volume} {40}},\ \bibinfo {pages} {279} (\bibinfo {year}
  {1978})}\BibitemShut {NoStop}%
\bibitem [{\citenamefont {Graham}\ \emph
  {et~al.}(2015{\natexlab{a}})\citenamefont {Graham}, \citenamefont {Kaplan},\
  and\ \citenamefont {Rajendran}}]{Graham:2015cka}%
  \BibitemOpen
  \bibfield  {author} {\bibinfo {author} {\bibfnamefont {P.~W.}\ \bibnamefont
  {Graham}}, \bibinfo {author} {\bibfnamefont {D.~E.}\ \bibnamefont {Kaplan}},\
  and\ \bibinfo {author} {\bibfnamefont {S.}~\bibnamefont {Rajendran}},\
  }\bibfield  {title} {\bibinfo {title} {"{Cosmological Relaxation of the
  Electroweak Scale}"},\ }\href@noop {} {\bibfield  {journal} {\bibinfo
  {journal} {Phys. Rev. Lett.}\ }\textbf {\bibinfo {volume} {115}},\ \bibinfo
  {pages} {221801} (\bibinfo {year} {2015}{\natexlab{a}})}\BibitemShut
  {NoStop}%
\bibitem [{\citenamefont {Nomura}\ and\ \citenamefont
  {Thaler}(2009)}]{Nomura:2008ru}%
  \BibitemOpen
  \bibfield  {author} {\bibinfo {author} {\bibfnamefont {Y.}~\bibnamefont
  {Nomura}}\ and\ \bibinfo {author} {\bibfnamefont {J.}~\bibnamefont
  {Thaler}},\ }\bibfield  {title} {\bibinfo {title} {"{Dark Matter through the
  Axion Portal}"},\ }\href@noop {} {\bibfield  {journal} {\bibinfo  {journal}
  {Phys. Rev. D}\ }\textbf {\bibinfo {volume} {79}},\ \bibinfo {pages} {075008}
  (\bibinfo {year} {2009})}\BibitemShut {NoStop}%
\bibitem [{\citenamefont {Freytsis}\ and\ \citenamefont
  {Ligeti}(2011)}]{Freytsis:2010ne}%
  \BibitemOpen
  \bibfield  {author} {\bibinfo {author} {\bibfnamefont {M.}~\bibnamefont
  {Freytsis}}\ and\ \bibinfo {author} {\bibfnamefont {Z.}~\bibnamefont
  {Ligeti}},\ }\bibfield  {title} {\bibinfo {title} {{On dark matter models
  with uniquely spin-dependent detection possibilities}},\ }\href@noop {}
  {\bibfield  {journal} {\bibinfo  {journal} {Phys. Rev. D}\ }\textbf {\bibinfo
  {volume} {83}},\ \bibinfo {pages} {115009} (\bibinfo {year}
  {2011})}\BibitemShut {NoStop}%
\bibitem [{\citenamefont {Dolan}\ \emph {et~al.}(2015)\citenamefont {Dolan},
  \citenamefont {Kahlhoefer}, \citenamefont {McCabe},\ and\ \citenamefont
  {Schmidt-Hoberg}}]{Dolan:2014ska}%
  \BibitemOpen
  \bibfield  {author} {\bibinfo {author} {\bibfnamefont {M.~J.}\ \bibnamefont
  {Dolan}}, \bibinfo {author} {\bibfnamefont {F.}~\bibnamefont {Kahlhoefer}},
  \bibinfo {author} {\bibfnamefont {C.}~\bibnamefont {McCabe}},\ and\ \bibinfo
  {author} {\bibfnamefont {K.}~\bibnamefont {Schmidt-Hoberg}},\ }\bibfield
  {title} {\bibinfo {title} {"{A taste of dark matter: Flavour constraints on
  pseudoscalar mediators}"},\ }\href@noop {} {\bibfield  {journal} {\bibinfo
  {journal} {JHEP}\ }\textbf {\bibinfo {volume} {03}},\ \bibinfo {pages}
  {171}}\BibitemShut {NoStop}%
\bibitem [{\citenamefont {Hochberg}\ \emph {et~al.}(2018)\citenamefont
  {Hochberg}, \citenamefont {Kuflik}, \citenamefont {Mcgehee}, \citenamefont
  {Murayama},\ and\ \citenamefont {Schutz}}]{Hochberg:2018rjs}%
  \BibitemOpen
  \bibfield  {author} {\bibinfo {author} {\bibfnamefont {Y.}~\bibnamefont
  {Hochberg}}, \bibinfo {author} {\bibfnamefont {E.}~\bibnamefont {Kuflik}},
  \bibinfo {author} {\bibfnamefont {R.}~\bibnamefont {Mcgehee}}, \bibinfo
  {author} {\bibfnamefont {H.}~\bibnamefont {Murayama}},\ and\ \bibinfo
  {author} {\bibfnamefont {K.}~\bibnamefont {Schutz}},\ }\bibfield  {title}
  {\bibinfo {title} {{Strongly interacting massive particles through the axion
  portal}},\ }\href@noop {} {\bibfield  {journal} {\bibinfo  {journal} {Phys.
  Rev. D}\ }\textbf {\bibinfo {volume} {98}},\ \bibinfo {pages} {115031}
  (\bibinfo {year} {2018})}\BibitemShut {NoStop}%
\bibitem [{\citenamefont {Marsh}(2016)}]{Marsh:2015xka}%
  \BibitemOpen
  \bibfield  {author} {\bibinfo {author} {\bibfnamefont {D.~J.~E.}\
  \bibnamefont {Marsh}},\ }\bibfield  {title} {\bibinfo {title} {"{Axion
  Cosmology}"},\ }\href@noop {} {\bibfield  {journal} {\bibinfo  {journal}
  {Phys. Rept.}\ }\textbf {\bibinfo {volume} {643}},\ \bibinfo {pages} {1}
  (\bibinfo {year} {2016})}\BibitemShut {NoStop}%
\bibitem [{\citenamefont {Graham}\ \emph
  {et~al.}(2015{\natexlab{b}})\citenamefont {Graham}, \citenamefont
  {Irastorza}, \citenamefont {Lamoreaux}, \citenamefont {Lindner},\ and\
  \citenamefont {van Bibber}}]{Graham:2015ouw}%
  \BibitemOpen
  \bibfield  {author} {\bibinfo {author} {\bibfnamefont {P.~W.}\ \bibnamefont
  {Graham}}, \bibinfo {author} {\bibfnamefont {I.~G.}\ \bibnamefont
  {Irastorza}}, \bibinfo {author} {\bibfnamefont {S.~K.}\ \bibnamefont
  {Lamoreaux}}, \bibinfo {author} {\bibfnamefont {A.}~\bibnamefont {Lindner}},\
  and\ \bibinfo {author} {\bibfnamefont {K.~A.}\ \bibnamefont {van Bibber}},\
  }\bibfield  {title} {\bibinfo {title} {"{Experimental Searches for the Axion
  and Axion-Like Particles}"},\ }\href@noop {} {\bibfield  {journal} {\bibinfo
  {journal} {Ann. Rev. Nucl. Part. Sci.}\ }\textbf {\bibinfo {volume} {65}},\
  \bibinfo {pages} {485} (\bibinfo {year} {2015}{\natexlab{b}})}\BibitemShut
  {NoStop}%
\bibitem [{\citenamefont {Hook}(2019)}]{Hook:2018dlk}%
  \BibitemOpen
  \bibfield  {author} {\bibinfo {author} {\bibfnamefont {A.}~\bibnamefont
  {Hook}},\ }\bibfield  {title} {\bibinfo {title} {"{TASI Lectures on the
  Strong CP Problem and Axions}"},\ }\href@noop {} {\bibfield  {journal}
  {\bibinfo  {journal} {PoS}\ }\textbf {\bibinfo {volume} {TASI2018}},\
  \bibinfo {pages} {004} (\bibinfo {year} {2019})}\BibitemShut {NoStop}%
\bibitem [{\citenamefont {Irastorza}\ and\ \citenamefont
  {Redondo}(2018)}]{Irastorza:2018dyq}%
  \BibitemOpen
  \bibfield  {author} {\bibinfo {author} {\bibfnamefont {I.~G.}\ \bibnamefont
  {Irastorza}}\ and\ \bibinfo {author} {\bibfnamefont {J.}~\bibnamefont
  {Redondo}},\ }\bibfield  {title} {\bibinfo {title} {{New experimental
  approaches in the search for axion-like particles}},\ }\href@noop {}
  {\bibfield  {journal} {\bibinfo  {journal} {Prog. Part. Nucl. Phys.}\
  }\textbf {\bibinfo {volume} {102}},\ \bibinfo {pages} {89} (\bibinfo {year}
  {2018})}\BibitemShut {NoStop}%
\bibitem [{\citenamefont {Choi}\ \emph {et~al.}(2021)\citenamefont {Choi},
  \citenamefont {Im},\ and\ \citenamefont {Sub~Shin}}]{Choi:2020rgn}%
  \BibitemOpen
  \bibfield  {author} {\bibinfo {author} {\bibfnamefont {K.}~\bibnamefont
  {Choi}}, \bibinfo {author} {\bibfnamefont {S.~H.}\ \bibnamefont {Im}},\ and\
  \bibinfo {author} {\bibfnamefont {C.}~\bibnamefont {Sub~Shin}},\ }\bibfield
  {title} {\bibinfo {title} {"{Recent Progress in the Physics of Axions and
  Axion-Like Particles}"},\ }\href@noop {} {\bibfield  {journal} {\bibinfo
  {journal} {Ann. Rev. Nucl. Part. Sci.}\ }\textbf {\bibinfo {volume} {71}},\
  \bibinfo {pages} {225} (\bibinfo {year} {2021})}\BibitemShut {NoStop}%
\bibitem [{\citenamefont {Bauer}\ \emph {et~al.}(2022)\citenamefont {Bauer},
  \citenamefont {Neubert}, \citenamefont {Renner}, \citenamefont {Schnubel},\
  and\ \citenamefont {Thamm}}]{Bauer:2021mvw}%
  \BibitemOpen
  \bibfield  {author} {\bibinfo {author} {\bibfnamefont {M.}~\bibnamefont
  {Bauer}}, \bibinfo {author} {\bibfnamefont {M.}~\bibnamefont {Neubert}},
  \bibinfo {author} {\bibfnamefont {S.}~\bibnamefont {Renner}}, \bibinfo
  {author} {\bibfnamefont {M.}~\bibnamefont {Schnubel}},\ and\ \bibinfo
  {author} {\bibfnamefont {A.}~\bibnamefont {Thamm}},\ }\bibfield  {title}
  {\bibinfo {title} {{Flavor probes of axion-like particles}},\ }\href@noop {}
  {\bibfield  {journal} {\bibinfo  {journal} {JHEP}\ }\textbf {\bibinfo
  {volume} {09}},\ \bibinfo {pages} {056}}\BibitemShut {NoStop}%
\bibitem [{\citenamefont {Jerhot}\ \emph {et~al.}(2022)\citenamefont {Jerhot},
  \citenamefont {D\"obrich}, \citenamefont {Ertas}, \citenamefont
  {Kahlhoefer},\ and\ \citenamefont {Spadaro}}]{Jerhot:2022chi}%
  \BibitemOpen
  \bibfield  {author} {\bibinfo {author} {\bibfnamefont {J.}~\bibnamefont
  {Jerhot}}, \bibinfo {author} {\bibfnamefont {B.}~\bibnamefont {D\"obrich}},
  \bibinfo {author} {\bibfnamefont {F.}~\bibnamefont {Ertas}}, \bibinfo
  {author} {\bibfnamefont {F.}~\bibnamefont {Kahlhoefer}},\ and\ \bibinfo
  {author} {\bibfnamefont {T.}~\bibnamefont {Spadaro}},\ }\bibfield  {title}
  {\bibinfo {title} {"{ALPINIST: Axion-Like Particles In Numerous Interactions
  Simulated and Tabulated}"},\ }\href@noop {} {\bibfield  {journal} {\bibinfo
  {journal} {JHEP}\ }\textbf {\bibinfo {volume} {07}},\ \bibinfo {pages}
  {094}}\BibitemShut {NoStop}%
\bibitem [{\citenamefont {Lanfranchi}\ \emph {et~al.}(2021)\citenamefont
  {Lanfranchi}, \citenamefont {Pospelov},\ and\ \citenamefont
  {Schuster}}]{Lanfranchi:2020crw}%
  \BibitemOpen
  \bibfield  {author} {\bibinfo {author} {\bibfnamefont {G.}~\bibnamefont
  {Lanfranchi}}, \bibinfo {author} {\bibfnamefont {M.}~\bibnamefont
  {Pospelov}},\ and\ \bibinfo {author} {\bibfnamefont {P.}~\bibnamefont
  {Schuster}},\ }\bibfield  {title} {\bibinfo {title} {"{The Search for Feebly
  Interacting Particles}"},\ }\href@noop {} {\bibfield  {journal} {\bibinfo
  {journal} {Ann. Rev. Nucl. Part. Sci.}\ }\textbf {\bibinfo {volume} {71}},\
  \bibinfo {pages} {279} (\bibinfo {year} {2021})}\BibitemShut {NoStop}%
\bibitem [{\citenamefont {Antel}\ \emph {et~al.}(2023)\citenamefont {Antel}
  \emph {et~al.}}]{Antel:2023hkf}%
  \BibitemOpen
  \bibfield  {author} {\bibinfo {author} {\bibfnamefont {C.}~\bibnamefont
  {Antel}} \emph {et~al.},\ }\bibfield  {title} {\bibinfo {title} {"{Feebly
  Interacting Particles: FIPs 2022 workshop report}"},\ }in\ \href@noop {}
  {\emph {\bibinfo {booktitle} {{Workshop on Feebly-Interacting Particles}}}}\
  (\bibinfo {year} {2023})\BibitemShut {NoStop}%
\bibitem [{\citenamefont {Agrawal}\ \emph {et~al.}(2021)\citenamefont {Agrawal}
  \emph {et~al.}}]{Agrawal:2021dbo}%
  \BibitemOpen
  \bibfield  {author} {\bibinfo {author} {\bibfnamefont {P.}~\bibnamefont
  {Agrawal}} \emph {et~al.},\ }\bibfield  {title} {\bibinfo {title}
  {"{Feebly-interacting particles: FIPs 2020 workshop report}"},\ }\href@noop
  {} {\bibfield  {journal} {\bibinfo  {journal} {Eur. Phys. J. C}\ }\textbf
  {\bibinfo {volume} {81}},\ \bibinfo {pages} {1015} (\bibinfo {year}
  {2021})}\BibitemShut {NoStop}%
\bibitem [{\citenamefont {Aloni}\ \emph {et~al.}(2019)\citenamefont {Aloni},
  \citenamefont {Fanelli}, \citenamefont {Soreq},\ and\ \citenamefont
  {Williams}}]{Aloni:2019ruo}%
  \BibitemOpen
  \bibfield  {author} {\bibinfo {author} {\bibfnamefont {D.}~\bibnamefont
  {Aloni}}, \bibinfo {author} {\bibfnamefont {C.}~\bibnamefont {Fanelli}},
  \bibinfo {author} {\bibfnamefont {Y.}~\bibnamefont {Soreq}},\ and\ \bibinfo
  {author} {\bibfnamefont {M.}~\bibnamefont {Williams}},\ }\bibfield  {title}
  {\bibinfo {title} {"{Photoproduction of Axionlike Particles}"},\ }\href@noop
  {} {\bibfield  {journal} {\bibinfo  {journal} {Phys. Rev. Lett.}\ }\textbf
  {\bibinfo {volume} {123}},\ \bibinfo {pages} {071801} (\bibinfo {year}
  {2019})}\BibitemShut {NoStop}%
\bibitem [{\citenamefont {Primakoff}(1951)}]{PhysRev.81.899}%
  \BibitemOpen
  \bibfield  {author} {\bibinfo {author} {\bibfnamefont {H.}~\bibnamefont
  {Primakoff}},\ }\bibfield  {title} {\bibinfo {title} {Photo-production of
  neutral mesons in nuclear electric fields and the mean life of the neutral
  meson},\ }\href {https://doi.org/10.1103/PhysRev.81.899} {\bibfield
  {journal} {\bibinfo  {journal} {Phys. Rev.}\ }\textbf {\bibinfo {volume}
  {81}},\ \bibinfo {pages} {899} (\bibinfo {year} {1951})}\BibitemShut
  {NoStop}%
\bibitem [{\citenamefont {Adhikari}\ \emph {et~al.}(2022)\citenamefont
  {Adhikari} \emph {et~al.}}]{Adhikari_2022}%
  \BibitemOpen
  \bibfield  {author} {\bibinfo {author} {\bibfnamefont {S.}~\bibnamefont
  {Adhikari}} \emph {et~al.},\ }\bibfield  {title} {\bibinfo {title} {Search
  for photoproduction of axionlike particles at {GlueX}},\ }\bibfield
  {journal} {\bibinfo  {journal} {Physical Review D}\ }\textbf {\bibinfo
  {volume} {105}},\ \href {https://doi.org/10.1103/physrevd.105.052007}
  {10.1103/physrevd.105.052007} (\bibinfo {year} {2022})\BibitemShut {NoStop}%
\bibitem [{\citenamefont {{Hen, O. and others}}()}]{E12:11:003A}%
  \BibitemOpen
  \bibfield  {author} {\bibinfo {author} {\bibnamefont {{Hen, O. and
  others}}},\ }\href@noop {} {\bibinfo {title} {{Jefferson Lab 12 GeV
  experiment E12-11-003A}}},\ \bibinfo {howpublished}
  {https://www.jlab.org/exp\_prog/proposals/15/E12-11-003A.pdf}\BibitemShut
  {NoStop}%
\bibitem [{\citenamefont {Hen}\ \emph {et~al.}(2017)\citenamefont {Hen},
  \citenamefont {Miller}, \citenamefont {Piasetzky},\ and\ \citenamefont
  {Weinstein}}]{RevModPhys.89.045002}%
  \BibitemOpen
  \bibfield  {author} {\bibinfo {author} {\bibfnamefont {O.}~\bibnamefont
  {Hen}}, \bibinfo {author} {\bibfnamefont {G.~A.}\ \bibnamefont {Miller}},
  \bibinfo {author} {\bibfnamefont {E.}~\bibnamefont {Piasetzky}},\ and\
  \bibinfo {author} {\bibfnamefont {L.~B.}\ \bibnamefont {Weinstein}},\
  }\bibfield  {title} {\bibinfo {title} {Nucleon-nucleon correlations,
  short-lived excitations, and the quarks within},\ }\href
  {https://doi.org/10.1103/RevModPhys.89.045002} {\bibfield  {journal}
  {\bibinfo  {journal} {Rev. Mod. Phys.}\ }\textbf {\bibinfo {volume} {89}},\
  \bibinfo {pages} {045002} (\bibinfo {year} {2017})}\BibitemShut {NoStop}%
\bibitem [{\citenamefont {Dutta}\ \emph {et~al.}(2013)\citenamefont {Dutta},
  \citenamefont {Hafidi},\ and\ \citenamefont {Strikman}}]{Dutta_2013}%
  \BibitemOpen
  \bibfield  {author} {\bibinfo {author} {\bibfnamefont {D.}~\bibnamefont
  {Dutta}}, \bibinfo {author} {\bibfnamefont {K.}~\bibnamefont {Hafidi}},\ and\
  \bibinfo {author} {\bibfnamefont {M.}~\bibnamefont {Strikman}},\ }\bibfield
  {title} {\bibinfo {title} {Color transparency: Past, present and future},\
  }\href {https://doi.org/10.1016/j.ppnp.2012.11.001} {\bibfield  {journal}
  {\bibinfo  {journal} {Progress in Particle and Nuclear Physics}\ }\textbf
  {\bibinfo {volume} {69}},\ \bibinfo {pages} {1} (\bibinfo {year}
  {2013})}\BibitemShut {NoStop}%
\bibitem [{\citenamefont {Leemann}\ \emph {et~al.}(2001)\citenamefont
  {Leemann}, \citenamefont {Douglas},\ and\ \citenamefont
  {Krafft}}]{doi:10.1146/annurev.nucl.51.101701.132327}%
  \BibitemOpen
  \bibfield  {author} {\bibinfo {author} {\bibfnamefont {C.~W.}\ \bibnamefont
  {Leemann}}, \bibinfo {author} {\bibfnamefont {D.~R.}\ \bibnamefont
  {Douglas}},\ and\ \bibinfo {author} {\bibfnamefont {G.~A.}\ \bibnamefont
  {Krafft}},\ }\bibfield  {title} {\bibinfo {title} {{THE CONTINUOUS ELECTRON
  BEAM ACCELERATOR FACILITY: CEBAF at the Jefferson Laboratory}},\ }\href
  {https://doi.org/10.1146/annurev.nucl.51.101701.132327} {\bibfield  {journal}
  {\bibinfo  {journal} {Annual Review of Nuclear and Particle Science}\
  }\textbf {\bibinfo {volume} {51}},\ \bibinfo {pages} {413} (\bibinfo {year}
  {2001})},\ \Eprint
  {https://arxiv.org/abs/https://doi.org/10.1146/annurev.nucl.51.101701.132327}
  {https://doi.org/10.1146/annurev.nucl.51.101701.132327} \BibitemShut
  {NoStop}%
\bibitem [{\citenamefont {Barbosa}\ \emph
  {et~al.}(2015{\natexlab{a}})\citenamefont {Barbosa}, \citenamefont {Hutton},
  \citenamefont {Sitnikov}, \citenamefont {Somov}, \citenamefont {Somov},\ and\
  \citenamefont {Tolstukhin}}]{Hodoscope}%
  \BibitemOpen
  \bibfield  {author} {\bibinfo {author} {\bibfnamefont {F.}~\bibnamefont
  {Barbosa}}, \bibinfo {author} {\bibfnamefont {C.}~\bibnamefont {Hutton}},
  \bibinfo {author} {\bibfnamefont {A.}~\bibnamefont {Sitnikov}}, \bibinfo
  {author} {\bibfnamefont {A.}~\bibnamefont {Somov}}, \bibinfo {author}
  {\bibfnamefont {S.}~\bibnamefont {Somov}},\ and\ \bibinfo {author}
  {\bibfnamefont {I.}~\bibnamefont {Tolstukhin}},\ }\bibfield  {title}
  {\bibinfo {title} {{Pair spectrometer hodoscope for Hall D at Jefferson
  Lab}},\ }\href@noop {} {\bibfield  {journal} {\bibinfo  {journal} {Nuclear
  Instruments and Methods in Physics Research Section A: Accelerators,
  Spectrometers, Detectors and Associated Equipment}\ }\textbf {\bibinfo
  {volume} {795}},\ \bibinfo {pages} {376} (\bibinfo {year}
  {2015}{\natexlab{a}})}\BibitemShut {NoStop}%
\bibitem [{\citenamefont {Adhikari}\ \emph {et~al.}(2021)\citenamefont
  {Adhikari} \emph {et~al.}}]{GlueXNIM}%
  \BibitemOpen
  \bibfield  {author} {\bibinfo {author} {\bibfnamefont {S.}~\bibnamefont
  {Adhikari}} \emph {et~al.},\ }\bibfield  {title} {\bibinfo {title} {{The
  GlueX beamline and detector}},\ }\href
  {https://doi.org/https://doi.org/10.1016/j.nima.2020.164807} {\bibfield
  {journal} {\bibinfo  {journal} {Nuclear Instruments and Methods in Physics
  Research Section A: Accelerators, Spectrometers, Detectors and Associated
  Equipment}\ }\textbf {\bibinfo {volume} {987}},\ \bibinfo {pages} {164807}
  (\bibinfo {year} {2021})}\BibitemShut {NoStop}%
\bibitem [{\citenamefont {Pooser}\ \emph {et~al.}(2019)\citenamefont {Pooser}
  \emph {et~al.}}]{SC}%
  \BibitemOpen
  \bibfield  {author} {\bibinfo {author} {\bibfnamefont {E.}~\bibnamefont
  {Pooser}} \emph {et~al.},\ }\bibfield  {title} {\bibinfo {title} {{The GlueX
  Start Counter Detector}},\ }\href@noop {} {\bibfield  {journal} {\bibinfo
  {journal} {Nuclear Instruments and Methods in Physics Research Section A:
  Accelerators, Spectrometers, Detectors and Associated Equipment}\ }\textbf
  {\bibinfo {volume} {927}},\ \bibinfo {pages} {330} (\bibinfo {year}
  {2019})}\BibitemShut {NoStop}%
\bibitem [{\citenamefont {Jarvis}\ \emph {et~al.}(2020)\citenamefont {Jarvis}
  \emph {et~al.}}]{CDC}%
  \BibitemOpen
  \bibfield  {author} {\bibinfo {author} {\bibfnamefont {N.}~\bibnamefont
  {Jarvis}} \emph {et~al.},\ }\bibfield  {title} {\bibinfo {title} {{The
  Central Drift Chamber for GlueX}},\ }\href@noop {} {\bibfield  {journal}
  {\bibinfo  {journal} {Nuclear Instruments and Methods in Physics Research
  Section A: Accelerators, Spectrometers, Detectors and Associated Equipment}\
  }\textbf {\bibinfo {volume} {962}},\ \bibinfo {pages} {163727} (\bibinfo
  {year} {2020})}\BibitemShut {NoStop}%
\bibitem [{\citenamefont {Beattie}\ \emph {et~al.}(2018)\citenamefont {Beattie}
  \emph {et~al.}}]{BCAL}%
  \BibitemOpen
  \bibfield  {author} {\bibinfo {author} {\bibfnamefont {T.}~\bibnamefont
  {Beattie}} \emph {et~al.},\ }\bibfield  {title} {\bibinfo {title}
  {{Construction and performance of the barrel electromagnetic calorimeter for
  the GlueX experiment}},\ }\href@noop {} {\bibfield  {journal} {\bibinfo
  {journal} {Nuclear Instruments and Methods in Physics Research Section A:
  Accelerators, Spectrometers, Detectors and Associated Equipment}\ }\textbf
  {\bibinfo {volume} {896}},\ \bibinfo {pages} {24} (\bibinfo {year}
  {2018})}\BibitemShut {NoStop}%
\bibitem [{\citenamefont {Pentchev}\ \emph {et~al.}(2017)\citenamefont
  {Pentchev}, \citenamefont {Barbosa}, \citenamefont {Berdnikov}, \citenamefont
  {Butler}, \citenamefont {Furletov}, \citenamefont {Robison},\ and\
  \citenamefont {Zihlmann}}]{FDC}%
  \BibitemOpen
  \bibfield  {author} {\bibinfo {author} {\bibfnamefont {L.}~\bibnamefont
  {Pentchev}}, \bibinfo {author} {\bibfnamefont {F.}~\bibnamefont {Barbosa}},
  \bibinfo {author} {\bibfnamefont {V.}~\bibnamefont {Berdnikov}}, \bibinfo
  {author} {\bibfnamefont {D.}~\bibnamefont {Butler}}, \bibinfo {author}
  {\bibfnamefont {S.}~\bibnamefont {Furletov}}, \bibinfo {author}
  {\bibfnamefont {L.}~\bibnamefont {Robison}},\ and\ \bibinfo {author}
  {\bibfnamefont {B.}~\bibnamefont {Zihlmann}},\ }\bibfield  {title} {\bibinfo
  {title} {{Studies with cathode drift chambers for the GlueX experiment at
  Jefferson Lab}},\ }\href@noop {} {\bibfield  {journal} {\bibinfo  {journal}
  {{Nuclear Instruments and Methods in Physics Research Section A:
  Accelerators, Spectrometers, Detectors and Associated Equipment}}\ }\textbf
  {\bibinfo {volume} {845}},\ \bibinfo {pages} {281} (\bibinfo {year}
  {2017})}\BibitemShut {NoStop}%
\bibitem [{\citenamefont {Moriya}\ \emph {et~al.}(2013)\citenamefont {Moriya}
  \emph {et~al.}}]{FCAL}%
  \BibitemOpen
  \bibfield  {author} {\bibinfo {author} {\bibfnamefont {K.}~\bibnamefont
  {Moriya}} \emph {et~al.},\ }\bibfield  {title} {\bibinfo {title} {{A
  measurement of the energy and timing resolution of the GlueX Forward
  Calorimeter using an electron beam}},\ }\href@noop {} {\bibfield  {journal}
  {\bibinfo  {journal} {Nuclear Instruments and Methods in Physics Research
  Section A: Accelerators, Spectrometers, Detectors and Associated Equipment}\
  }\textbf {\bibinfo {volume} {726}},\ \bibinfo {pages} {60} (\bibinfo {year}
  {2013})}\BibitemShut {NoStop}%
\bibitem [{\citenamefont {Cowan}\ \emph {et~al.}(2011)\citenamefont {Cowan},
  \citenamefont {Cranmer}, \citenamefont {Gross},\ and\ \citenamefont
  {Vitells}}]{Cowan2013}%
  \BibitemOpen
  \bibfield  {author} {\bibinfo {author} {\bibfnamefont {G.}~\bibnamefont
  {Cowan}}, \bibinfo {author} {\bibfnamefont {K.}~\bibnamefont {Cranmer}},
  \bibinfo {author} {\bibfnamefont {E.}~\bibnamefont {Gross}},\ and\ \bibinfo
  {author} {\bibfnamefont {O.}~\bibnamefont {Vitells}},\ }\bibfield  {title}
  {\bibinfo {title} {Asymptotic formulae for likelihood-based tests of new
  physics},\ }\href {https://doi.org/10.1140/epjc/s10052-011-1554-0} {\bibfield
   {journal} {\bibinfo  {journal} {The European Physical Journal C}\ }\textbf
  {\bibinfo {volume} {71}},\ \bibinfo {pages} {1554} (\bibinfo {year}
  {2011})}\BibitemShut {NoStop}%
\bibitem [{\citenamefont {Workman}\ and\ \citenamefont {Others}(2022)}]{PDG}%
  \BibitemOpen
  \bibfield  {author} {\bibinfo {author} {\bibfnamefont {R.~L.}\ \bibnamefont
  {Workman}}\ and\ \bibinfo {author} {\bibnamefont {Others}} (\bibinfo
  {collaboration} {Particle Data Group}),\ }\bibfield  {title} {\bibinfo
  {title} {{Review of Particle Physics}},\ }\href
  {https://doi.org/10.1093/ptep/ptac097} {\bibfield  {journal} {\bibinfo
  {journal} {PTEP}\ }\textbf {\bibinfo {volume} {2022}},\ \bibinfo {pages}
  {083C01} (\bibinfo {year} {2022})}\BibitemShut {NoStop}%
\bibitem [{\citenamefont {Gevorkyan}\ \emph {et~al.}(2009)\citenamefont
  {Gevorkyan}, \citenamefont {Gasparian}, \citenamefont {Gan}, \citenamefont
  {Larin},\ and\ \citenamefont {Khandaker}}]{PhysRevC.80.055201}%
  \BibitemOpen
  \bibfield  {author} {\bibinfo {author} {\bibfnamefont {S.}~\bibnamefont
  {Gevorkyan}}, \bibinfo {author} {\bibfnamefont {A.}~\bibnamefont
  {Gasparian}}, \bibinfo {author} {\bibfnamefont {L.}~\bibnamefont {Gan}},
  \bibinfo {author} {\bibfnamefont {I.}~\bibnamefont {Larin}},\ and\ \bibinfo
  {author} {\bibfnamefont {M.}~\bibnamefont {Khandaker}},\ }\bibfield  {title}
  {\bibinfo {title} {Photoproduction of pseudoscalar mesons off nuclei at
  forward angles},\ }\href {https://doi.org/10.1103/PhysRevC.80.055201}
  {\bibfield  {journal} {\bibinfo  {journal} {Phys. Rev. C}\ }\textbf {\bibinfo
  {volume} {80}},\ \bibinfo {pages} {055201} (\bibinfo {year}
  {2009})}\BibitemShut {NoStop}%
\bibitem [{\citenamefont {Barbosa}\ \emph
  {et~al.}(2015{\natexlab{b}})\citenamefont {Barbosa}, \citenamefont {Hutton},
  \citenamefont {Sitnikov}, \citenamefont {Somov}, \citenamefont {Somov},\ and\
  \citenamefont {Tolstukhin}}]{BARBOSA2015376}%
  \BibitemOpen
  \bibfield  {author} {\bibinfo {author} {\bibfnamefont {F.}~\bibnamefont
  {Barbosa}}, \bibinfo {author} {\bibfnamefont {C.}~\bibnamefont {Hutton}},
  \bibinfo {author} {\bibfnamefont {A.}~\bibnamefont {Sitnikov}}, \bibinfo
  {author} {\bibfnamefont {A.}~\bibnamefont {Somov}}, \bibinfo {author}
  {\bibfnamefont {S.}~\bibnamefont {Somov}},\ and\ \bibinfo {author}
  {\bibfnamefont {I.}~\bibnamefont {Tolstukhin}},\ }\bibfield  {title}
  {\bibinfo {title} {Pair spectrometer hodoscope for hall d at jefferson lab},\
  }\href@noop {} {\bibfield  {journal} {\bibinfo  {journal} {Nuclear
  Instruments and Methods in Physics Research Section A: Accelerators,
  Spectrometers, Detectors and Associated Equipment}\ }\textbf {\bibinfo
  {volume} {795}},\ \bibinfo {pages} {376} (\bibinfo {year}
  {2015}{\natexlab{b}})}\BibitemShut {NoStop}%
\bibitem [{\citenamefont {Bressler}\ \emph {et~al.}(2014)\citenamefont
  {Bressler}, \citenamefont {Dery},\ and\ \citenamefont
  {Efrati}}]{Bressler_2014}%
  \BibitemOpen
  \bibfield  {author} {\bibinfo {author} {\bibfnamefont {S.}~\bibnamefont
  {Bressler}}, \bibinfo {author} {\bibfnamefont {A.}~\bibnamefont {Dery}},\
  and\ \bibinfo {author} {\bibfnamefont {A.}~\bibnamefont {Efrati}},\
  }\bibfield  {title} {\bibinfo {title} {Asymmetric lepton-flavor violating
  higgs boson decays},\ }\bibfield  {journal} {\bibinfo  {journal} {Physical
  Review D}\ }\textbf {\bibinfo {volume} {90}},\ \href
  {https://doi.org/10.1103/physrevd.90.015025} {10.1103/physrevd.90.015025}
  (\bibinfo {year} {2014})\BibitemShut {NoStop}%
\bibitem [{\citenamefont {Aad}\ \emph {et~al.}(2017)\citenamefont {Aad} \emph
  {et~al.}}]{Aad_2017}%
  \BibitemOpen
  \bibfield  {author} {\bibinfo {author} {\bibfnamefont {G.}~\bibnamefont
  {Aad}} \emph {et~al.},\ }\bibfield  {title} {\bibinfo {title} {Search for
  lepton-flavour-violating decays of the higgs and z bosons with the {ATLAS}
  detector},\ }\bibfield  {journal} {\bibinfo  {journal} {The European Physical
  Journal C}\ }\textbf {\bibinfo {volume} {77}},\ \href
  {https://doi.org/10.1140/epjc/s10052-017-4624-0}
  {10.1140/epjc/s10052-017-4624-0} (\bibinfo {year} {2017})\BibitemShut
  {NoStop}%
\bibitem [{\citenamefont {Ablikim}\ \emph {et~al.}(2023)\citenamefont {Ablikim}
  \emph {et~al.}}]{BESIII:2022rzz}%
  \BibitemOpen
  \bibfield  {author} {\bibinfo {author} {\bibfnamefont {M.}~\bibnamefont
  {Ablikim}} \emph {et~al.} (\bibinfo {collaboration} {BESIII}),\ }\bibfield
  {title} {\bibinfo {title} {{Search for an axion-like particle in radiative
  J/\ensuremath{\psi} decays}},\ }\href
  {https://doi.org/10.1016/j.physletb.2023.137698} {\bibfield  {journal}
  {\bibinfo  {journal} {Phys. Lett. B}\ }\textbf {\bibinfo {volume} {838}},\
  \bibinfo {pages} {137698} (\bibinfo {year} {2023})},\ \Eprint
  {https://arxiv.org/abs/2211.12699} {arXiv:2211.12699 [hep-ex]} \BibitemShut
  {NoStop}%
\bibitem [{JEF()}]{JEF}%
  \BibitemOpen
  \href@noop {} {\bibinfo {title} {{An update on the GlueX II and Jefferson Lab
  $\eta$ Factory experiments}}},\ \bibinfo {howpublished}
  {https://www.jlab.org/exp\_prog/proposals/20/Jeopardy/E12-12-002\_Update.pdf}\BibitemShut
  {NoStop}%
\bibitem [{\citenamefont {Abbiendi~et al.}\ and\ \citenamefont
  {Collaboration}(2003)}]{OPAL}%
  \BibitemOpen
  \bibfield  {author} {\bibinfo {author} {\bibfnamefont {G.}~\bibnamefont
  {Abbiendi~et al.}}\ and\ \bibinfo {author} {\bibfnamefont {T.~O.}\
  \bibnamefont {Collaboration}},\ }\bibfield  {title} {\bibinfo {title}
  {{Multi-photon production in ee collisions at $\sqrt{s} = $181-209 GeV}},\
  }\href {https://doi.org/10.1140/epjc/s2002-01074-5} {\bibfield  {journal}
  {\bibinfo  {journal} {The European Physical Journal C - Particles and
  Fields}\ }\textbf {\bibinfo {volume} {26}},\ \bibinfo {pages} {331} (\bibinfo
  {year} {2003})}\BibitemShut {NoStop}%
\bibitem [{\citenamefont {Knapen}\ \emph {et~al.}(2017)\citenamefont {Knapen},
  \citenamefont {Lin}, \citenamefont {Lou},\ and\ \citenamefont
  {Melia}}]{PhysRevLett.118.171801}%
  \BibitemOpen
  \bibfield  {author} {\bibinfo {author} {\bibfnamefont {S.}~\bibnamefont
  {Knapen}}, \bibinfo {author} {\bibfnamefont {T.}~\bibnamefont {Lin}},
  \bibinfo {author} {\bibfnamefont {H.~K.}\ \bibnamefont {Lou}},\ and\ \bibinfo
  {author} {\bibfnamefont {T.}~\bibnamefont {Melia}},\ }\bibfield  {title}
  {\bibinfo {title} {{Searching for Axionlike Particles with Ultraperipheral
  Heavy-Ion Collisions}},\ }\href
  {https://doi.org/10.1103/PhysRevLett.118.171801} {\bibfield  {journal}
  {\bibinfo  {journal} {Phys. Rev. Lett.}\ }\textbf {\bibinfo {volume} {118}},\
  \bibinfo {pages} {171801} (\bibinfo {year} {2017})}\BibitemShut {NoStop}%
\bibitem [{\citenamefont {Bjorken}\ \emph {et~al.}(1988)\citenamefont
  {Bjorken}, \citenamefont {Ecklund}, \citenamefont {Nelson}, \citenamefont
  {Abashian}, \citenamefont {Church}, \citenamefont {Lu}, \citenamefont {Mo},
  \citenamefont {Nunamaker},\ and\ \citenamefont
  {Rassmann}}]{PhysRevD.38.3375}%
  \BibitemOpen
  \bibfield  {author} {\bibinfo {author} {\bibfnamefont {J.~D.}\ \bibnamefont
  {Bjorken}}, \bibinfo {author} {\bibfnamefont {S.}~\bibnamefont {Ecklund}},
  \bibinfo {author} {\bibfnamefont {W.~R.}\ \bibnamefont {Nelson}}, \bibinfo
  {author} {\bibfnamefont {A.}~\bibnamefont {Abashian}}, \bibinfo {author}
  {\bibfnamefont {C.}~\bibnamefont {Church}}, \bibinfo {author} {\bibfnamefont
  {B.}~\bibnamefont {Lu}}, \bibinfo {author} {\bibfnamefont {L.~W.}\
  \bibnamefont {Mo}}, \bibinfo {author} {\bibfnamefont {T.~A.}\ \bibnamefont
  {Nunamaker}},\ and\ \bibinfo {author} {\bibfnamefont {P.}~\bibnamefont
  {Rassmann}},\ }\bibfield  {title} {\bibinfo {title} {{Search for neutral
  metastable penetrating particles produced in the SLAC beam dump}},\ }\href
  {https://doi.org/10.1103/PhysRevD.38.3375} {\bibfield  {journal} {\bibinfo
  {journal} {Phys. Rev. D}\ }\textbf {\bibinfo {volume} {38}},\ \bibinfo
  {pages} {3375} (\bibinfo {year} {1988})}\BibitemShut {NoStop}%
\bibitem [{\citenamefont {Bl{\"u}mlein}\ \emph {et~al.}(1991)\citenamefont
  {Bl{\"u}mlein} \emph {et~al.}}]{BeamDumps}%
  \BibitemOpen
  \bibfield  {author} {\bibinfo {author} {\bibfnamefont {J.}~\bibnamefont
  {Bl{\"u}mlein}} \emph {et~al.},\ }\bibfield  {title} {\bibinfo {title}
  {Limits on neutral light scalar and pseudoscalar particles in a proton beam
  dump experiment},\ }\href {https://doi.org/10.1007/BF01548556} {\bibfield
  {journal} {\bibinfo  {journal} {Zeitschrift f{\"u}r Physik C Particles and
  Fields}\ }\textbf {\bibinfo {volume} {51}},\ \bibinfo {pages} {341} (\bibinfo
  {year} {1991})}\BibitemShut {NoStop}%
\bibitem [{\citenamefont {Mitridate}\ \emph {et~al.}(2023)\citenamefont
  {Mitridate}, \citenamefont {Papucci}, \citenamefont {Wang}, \citenamefont
  {Peña},\ and\ \citenamefont {Xie}}]{mitridate2023energetic}%
  \BibitemOpen
  \bibfield  {author} {\bibinfo {author} {\bibfnamefont {A.}~\bibnamefont
  {Mitridate}}, \bibinfo {author} {\bibfnamefont {M.}~\bibnamefont {Papucci}},
  \bibinfo {author} {\bibfnamefont {C.}~\bibnamefont {Wang}}, \bibinfo {author}
  {\bibfnamefont {C.}~\bibnamefont {Peña}},\ and\ \bibinfo {author}
  {\bibfnamefont {S.}~\bibnamefont {Xie}},\ }\href@noop {} {\bibinfo {title}
  {{Energetic long-lived particles in the CMS muon chambers}}} (\bibinfo {year}
  {2023}),\ \Eprint {https://arxiv.org/abs/2304.06109} {arXiv:2304.06109
  [hep-ph]} \BibitemShut {NoStop}%
\bibitem [{\citenamefont {D{\"o}brich}\ \emph {et~al.}(2016)\citenamefont
  {D{\"o}brich}, \citenamefont {Jaeckel}, \citenamefont {Kahlhoefer},
  \citenamefont {Ringwald},\ and\ \citenamefont {Schmidt-Hoberg}}]{SHiP}%
  \BibitemOpen
  \bibfield  {author} {\bibinfo {author} {\bibfnamefont {B.}~\bibnamefont
  {D{\"o}brich}}, \bibinfo {author} {\bibfnamefont {J.}~\bibnamefont
  {Jaeckel}}, \bibinfo {author} {\bibfnamefont {F.}~\bibnamefont {Kahlhoefer}},
  \bibinfo {author} {\bibfnamefont {A.}~\bibnamefont {Ringwald}},\ and\
  \bibinfo {author} {\bibfnamefont {K.}~\bibnamefont {Schmidt-Hoberg}},\
  }\bibfield  {title} {\bibinfo {title} {{ALPtraum: ALP production in proton
  beam dump experiments}},\ }\href {https://doi.org/10.1007/JHEP02(2016)018}
  {\bibfield  {journal} {\bibinfo  {journal} {Journal of High Energy Physics}\
  }\textbf {\bibinfo {volume} {2016}},\ \bibinfo {pages} {18} (\bibinfo {year}
  {2016})}\BibitemShut {NoStop}%
\bibitem [{\citenamefont {Feng}\ \emph {et~al.}(2018)\citenamefont {Feng},
  \citenamefont {Galon}, \citenamefont {Kling},\ and\ \citenamefont
  {Trojanowski}}]{PhysRevD.98.055021}%
  \BibitemOpen
  \bibfield  {author} {\bibinfo {author} {\bibfnamefont {J.~L.}\ \bibnamefont
  {Feng}}, \bibinfo {author} {\bibfnamefont {I.}~\bibnamefont {Galon}},
  \bibinfo {author} {\bibfnamefont {F.}~\bibnamefont {Kling}},\ and\ \bibinfo
  {author} {\bibfnamefont {S.}~\bibnamefont {Trojanowski}},\ }\bibfield
  {title} {\bibinfo {title} {{Axionlike particles at FASER: The LHC as a photon
  beam dump}},\ }\href {https://doi.org/10.1103/PhysRevD.98.055021} {\bibfield
  {journal} {\bibinfo  {journal} {Phys. Rev. D}\ }\textbf {\bibinfo {volume}
  {98}},\ \bibinfo {pages} {055021} (\bibinfo {year} {2018})}\BibitemShut
  {NoStop}%
\bibitem [{\citenamefont {Berlin}\ \emph {et~al.}(2018)\citenamefont {Berlin},
  \citenamefont {Gori}, \citenamefont {Schuster},\ and\ \citenamefont
  {Toro}}]{PhysRevD.98.035011}%
  \BibitemOpen
  \bibfield  {author} {\bibinfo {author} {\bibfnamefont {A.}~\bibnamefont
  {Berlin}}, \bibinfo {author} {\bibfnamefont {S.}~\bibnamefont {Gori}},
  \bibinfo {author} {\bibfnamefont {P.}~\bibnamefont {Schuster}},\ and\
  \bibinfo {author} {\bibfnamefont {N.}~\bibnamefont {Toro}},\ }\bibfield
  {title} {\bibinfo {title} {{Dark sectors at the Fermilab SeaQuest
  experiment}},\ }\href {https://doi.org/10.1103/PhysRevD.98.035011} {\bibfield
   {journal} {\bibinfo  {journal} {Phys. Rev. D}\ }\textbf {\bibinfo {volume}
  {98}},\ \bibinfo {pages} {035011} (\bibinfo {year} {2018})}\BibitemShut
  {NoStop}%
\bibitem [{\citenamefont {Dolan}\ \emph {et~al.}(2017)\citenamefont {Dolan},
  \citenamefont {Ferber}, \citenamefont {Hearty}, \citenamefont {Kahlhoefer},\
  and\ \citenamefont {Schmidt-Hoberg}}]{BelleII}%
  \BibitemOpen
  \bibfield  {author} {\bibinfo {author} {\bibfnamefont {M.~J.}\ \bibnamefont
  {Dolan}}, \bibinfo {author} {\bibfnamefont {T.}~\bibnamefont {Ferber}},
  \bibinfo {author} {\bibfnamefont {C.}~\bibnamefont {Hearty}}, \bibinfo
  {author} {\bibfnamefont {F.}~\bibnamefont {Kahlhoefer}},\ and\ \bibinfo
  {author} {\bibfnamefont {K.}~\bibnamefont {Schmidt-Hoberg}},\ }\bibfield
  {title} {\bibinfo {title} {{Revised constraints and Belle II sensitivity for
  visible and invisible axion-like particles}},\ }\href
  {https://doi.org/10.1007/JHEP12(2017)094} {\bibfield  {journal} {\bibinfo
  {journal} {Journal of High Energy Physics}\ }\textbf {\bibinfo {volume}
  {2017}},\ \bibinfo {pages} {94} (\bibinfo {year} {2017})}\BibitemShut
  {NoStop}%
\bibitem [{\citenamefont {Bai}\ \emph {et~al.}(2022)\citenamefont {Bai},
  \citenamefont {Blackburn}, \citenamefont {Borysov}, \citenamefont {Davidi},
  \citenamefont {Hartin}, \citenamefont {Heinemann}, \citenamefont {Ma},
  \citenamefont {Perez}, \citenamefont {Santra}, \citenamefont {Soreq},\ and\
  \citenamefont {Hod}}]{Bai_2022}%
  \BibitemOpen
  \bibfield  {author} {\bibinfo {author} {\bibfnamefont {Z.}~\bibnamefont
  {Bai}}, \bibinfo {author} {\bibfnamefont {T.}~\bibnamefont {Blackburn}},
  \bibinfo {author} {\bibfnamefont {O.}~\bibnamefont {Borysov}}, \bibinfo
  {author} {\bibfnamefont {O.}~\bibnamefont {Davidi}}, \bibinfo {author}
  {\bibfnamefont {A.}~\bibnamefont {Hartin}}, \bibinfo {author} {\bibfnamefont
  {B.}~\bibnamefont {Heinemann}}, \bibinfo {author} {\bibfnamefont
  {T.}~\bibnamefont {Ma}}, \bibinfo {author} {\bibfnamefont {G.}~\bibnamefont
  {Perez}}, \bibinfo {author} {\bibfnamefont {A.}~\bibnamefont {Santra}},
  \bibinfo {author} {\bibfnamefont {Y.}~\bibnamefont {Soreq}},\ and\ \bibinfo
  {author} {\bibfnamefont {N.~T.}\ \bibnamefont {Hod}},\ }\bibfield  {title}
  {\bibinfo {title} {New physics searches with an optical dump at {LUXE}},\
  }\bibfield  {journal} {\bibinfo  {journal} {Physical Review D}\ }\textbf
  {\bibinfo {volume} {106}},\ \href
  {https://doi.org/10.1103/physrevd.106.115034} {10.1103/physrevd.106.115034}
  (\bibinfo {year} {2022})\BibitemShut {NoStop}%
\bibitem [{\citenamefont {Abramowicz}\ \emph {et~al.}(2021)\citenamefont
  {Abramowicz} \emph {et~al.}}]{Abramowicz_2021}%
  \BibitemOpen
  \bibfield  {author} {\bibinfo {author} {\bibfnamefont {H.}~\bibnamefont
  {Abramowicz}} \emph {et~al.},\ }\bibfield  {title} {\bibinfo {title}
  {Conceptual design report for the {LUXE} experiment},\ }\href
  {https://doi.org/10.1140/epjs/s11734-021-00249-z} {\bibfield  {journal}
  {\bibinfo  {journal} {The European Physical Journal Special Topics}\ }\textbf
  {\bibinfo {volume} {230}},\ \bibinfo {pages} {2445} (\bibinfo {year}
  {2021})}\BibitemShut {NoStop}%
\end{thebibliography}%

\end{document}